\documentclass[aps,prd,preprint,groupedaddress,nofootinbib]{revtex4-1}

\usepackage{graphicx}
\usepackage{epsfig}
\usepackage{bm}
\usepackage{amsfonts}
\usepackage{amsmath}
\usepackage{amssymb}
\usepackage{multirow}
\usepackage{subfigure}
\usepackage{xcolor}
\graphicspath{{./fig/}}

\begin{document}



\title{The growth of DM and DE perturbations in DBI non-canonical scalar field scenario}





\author{K. Rezazadeh$^{1}$, S. Asadzadeh$^{1}$, K. Fahimi$^{1}$, K. Karami$^{1}$, and A. Mehrabi$^{2}$}

\affiliation{
$^{1}$Department of Physics, University of Kurdistan, Pasdaran Street, P.O. Box 66177-15175, Sanandaj, Iran\\
$^{2}$Department of Physics, Bu-Ali Sina University, Hamedan 65178, Iran
}


\date{\today}


\begin{abstract}

We study the effect of varying sound speed on clustering dark energy in the Dirac-Born-Infeld (DBI) scenario. The DBI action is included in the class of $k$-essence models, and it has an important role in describing the effective degrees of freedom of D-branes in the string theory. In the DBI setup, we take the anti-de Sitter (AdS) warp factor $f(\phi)=f_0\, \phi^{-4}$, and investigate the self-interacting quartic potential $V(\phi)=\lambda\phi^{4}/4$. We calculate the full expression of the effective sound speed for our model, and show that it can evolve with time during the cosmological evolution. Besides, the adiabatic sound speed evolves with time here, and this influences the background dynamics to some extent. We show that the effective sound speed is very close to the adiabatic sound speed. We examine the effect of the variable sound speed on growth of the perturbations in both the linear and non-linear regimes. In the linear regime, we apply the Pseudo-Newtonian formalism, and show that dark energy suppresses the growth of perturbations at low redshifts. From study the Integrated Sachs-Wolf (ISW) effect in our setup, we see that the model manifests some deviation from the concordance $\Lambda$CDM model. In the non-linear regime, we follow the approach of spherical collapse model, and calculate the linear overdensity $\delta_c(z_c)$, the virial overdensity $\Delta_{\rm vir}(z_c)$, overdensity at the turn around $\zeta(z_c)$ and the rate of expansion of collapsed region $h_{\rm ta}(z)$. Our results imply that the provided values of $\delta_c(z_c)$, $\Delta_{\rm vir}(z_c)$, $\zeta(z_c)$ and $h_{\rm ta}(z)$ in our clustering DBI dark energy model approach the fiducial value in the EdS universe at high enough redshifts. We further compute relative number density of halo objects above a given mass in our setting, and show that the number of structures with respect to the $\Lambda$CDM model is reduced more in the high mass tail at high redshifts.

\end{abstract}



\maketitle



\section{Introduction}
\label{section:introduction}

Cosmological data from different instruments indicate that the present-day Universe experiences an accelerating expansion \cite{Riess1998, Perlmutter1999, Wang1998, Tegmark2004, Alcaniz2004, Benjamin2007, Planck2018}. This acceleration is usually attributed to the dark energy (DE) whose nature is still unknown for us. The most familiar candidate for the DE is the Einstein cosmological constant $\Lambda$ which gives rise to a constant energy density for the Universe and this in turn leads to the negative pressure required for having an accelerating expansion. The cosmological model based on the cosmological constant is called the $\Lambda$CDM model, and it matches very well with observational data from the cosmic microwave background radiation and large-scale scale structure formation. In spite of its achievements, the cosmological constant suffers from some catastrophic drawbacks such as the fine-tuning and cosmic coincidence problems \cite{Weinberg1989, Sahni2000, Carroll2001, Padmanabhan2003, Copeland2006}. To overcome these problems, cosmologists consider the dynamical DE models. One important class of the dynamical DE models are based on a scalar field describing the evolution of vacuum energy content. In these kinds of models, the equation of state of the scalar field tends to the vacuum energy behavior just after the pressureless matter constituent which has dominated over radiation contribution. A cosmological solution which shows such a trend is called the tracker solution.

The most famous scalar field DE model is the quintessence model \cite{Caldwell1998, Erickson2002}, in which a scalar field with a canonical kinetic energy is coupled minimally to the Einstein gravity. The quintessence model can provide the tracker solutions required to solve the coincidence problem, but to do so, the parameters of the scalar potential should be adjusted very carefully. Therefore, the model requires high level of fine-tuning which is at odds with the primitive intentions of the model. An important alternative to the quintessence is the $k$-essence model \cite{Chiba2000, Armendariz-Picon2000, Armendariz-Picon2001} of DE, in which a generalized form is taken for the kinetic term of the scalar field. In contrast to the quintessence model, the tracker solutions of the $k$-essence model are general solutions of it, and hence the $k$-essence does not require high level of fine-tuning to overcome the coincidence problem. One privileged $k$-essence model which has robust theoretical motivations is the Dirac-Born-Infeld (DBI) dark energy model. The DBI action was formerly offered as a substitute for the standard action of electrodynamics. In the recent theoretical developments in the string theory, the DBI action has a prominence role in implying the effective D-brane degrees of freedom \cite{Bachas1996, Douglas1997, Burgess2003}.

So far, the DBI scalar field has been considered in different contexts of cosmology in the literature. Some of these articles have considered DBI as a source for inflation \cite{silverstein2004scalar, alishahiha2004dbi, chimento2008bridging, Amani:2018ueu, Rasouli:2018kvy} or dark energy \cite{Garousi:2004ph, martin2008dbi, ahn2010cosmological, Copeland:2010jt, Guo:2008sz, Panpanich:2017nft, kaeonikhom2012dynamics, Cai:2015rns, chimento2010dbi, Fahimi2018}. It has been shown in \cite{ahn2010cosmological} that the DBI model provides several new classes of dark energy behavior beyond quintessence due to its relativistic kinematics. In that paper, the authors discussed that the dark energy dynamics demonstrates attractor solutions which include the cosmological constant behavior. The authors also argued that the novel signature of DBI attractors is that the sound speed is driven to zero, unlike for quintessence where it is the speed of light. In \cite{Copeland:2010jt}, the dynamics of the DBI field is analyzed in a cosmological setup which includes a perfect fluid. There, the authors supposed arbitrary power law or exponential functions for the potential and the brane tension of the DBI field, and concluded that scaling solutions can exist if powers of the field in the potential and warp factor fulfill specific relations. The scaling solutions of the DBI scalar field also have been regarded in \cite{Guo:2008sz, Panpanich:2017nft}. The DBI dark energy field interacting with dark matter in terms of late-time scaling solutions was studied in \cite{kaeonikhom2012dynamics}. In \cite{chimento2010dbi} for unified DBI dark energy model, linear growth of perturbation was studied by fixing some of its degrees of freedom, and also a Bayesian analysis was performed to set observational constraints on the parameters of the model.

An attractive property of the DBI dark energy model is that the sound speed of the scalar field perturbations in this model can be different from the light speed. Therefore, it is expected that this feature has a noticeable implications in study of the large-scale structure formation in the Universe. In the process of structure formation, the gravitational instability causes the primordial density perturbation collapse \cite{Gunn1972, Press1974, White1978, Peebles2003, Ciardi2005, Bromm2011}. The primordial density perturbations are produced during the inflationary era \cite{Guth1981, Linde1990}. During inflation the Hubble horizon shrinks and so the wavelength of the cosmological perturbations exceeds the horizon size. In the subsequent stages of the Universe history, the horizon expands, and the perturbations enter the horizon again. At the early times after their reentry, the overdensities are small so that the linear theory of perturbations is applicable. In this period, the interesting scales in cosmology are much smaller than the size of the Hubble horizon, and also the velocities are non-relativistic. So, the linear regime of perturbations holds, and we can apply the Pseudo-Newtonian formalism to analyze the evolution of the overdensities. Since in the Pseudo-Newtonian formalism, the relativistic effects lead to the appearance of pressure terms in the Poisson equation, the Newtonian hydrodynamical equations can be used in the expanding Universe \cite{Abramo2008, Abramo2009JCAP}. However, at late times the perturbations grow and the overdensities enter the non-linear regime. From investigating the dynamics of the overdensities in the non-linear regime, we can obtain valuable cosmological predictions which can be assessed by observations. A simple analytical manner in study of the non-linear perturbations is the spherical collapse model (SCM) \cite{Gunn1972}. A feasibility of SCM is that in this scenario DE can behave like a fluid with clustering features same as those of DM \cite{Abramo2008}. This property of DE originates from the naive idea that when fluctuations in the fluid pressure grow, the effective equation of state of the collapsed sphere becomes different from that of the unperturbed background \cite{Abramo2008}. As discussed in \cite{Abramo2009JCAP}, the clustering properties of DE can manifest signals which are observable on the cosmological data.

In the present paper, we study the cosmological structure formation within the framework of DBI clustering DE. Although, we have investigated this subject previously in \cite{Fahimi2018} by assuming constant values for the adiabatic sound speed $c_s$ of the overdensities, but our investigating is more general and realistic here in several aspects. The most important one is that here we calculate the full expression of the effective sound speed $c_{\mathrm{eff}}$ and then use it in the perturbation equations. The effective sound speed is a key quantity which has a crucial role in the equations of the large-scale structure formation. The special importance of this parameter to some extent arises from the fact that it determines the amount of clustering DE. Despite this, calculation of this quantity is very complicated, and hence in most of the previous studies (see e.g. \cite{Batista2013, Pace2014, Malekjani2015, Malekjani:2016edh, Nazari-Pooya2016, Rezaei2017, Rezaei2017-2}), the effective sound speed either is set $c_{\mathrm{eff}}=1$, which is corresponding to the non-clustering case, or it is taken as $c_{\mathrm{eff}}=0$, which is related to the full-clustering case. In the present paper, we aim to provide a method to derive the full expression of the effective sound speed, which is applicable for all the $k$-essence DE models. In our DBI model, the effective sound speed can evolve with time, and thus our study for the cosmological structure formation will be more general since it is not limited to only the two special cases of non-clustering and full-clustering.

In addition to the effective sound speed $c_{\mathrm{eff}}$, the adiabatic sound speed $c_s$ can also evolve with time in our present work, and this is in contrast with our earlier work \cite{Fahimi2018} where $c_s$ was taken as a constant parameter. Evolution of $c_s$ during cosmological history influences the background dynamics, and hence it may lead to that the results at the background level which may differ from those of \cite{Fahimi2018}. In our setup, we can compare the effective sound speed $c_{\mathrm{eff}}$ and adiabatic sound speed $c_s$ during the cosmological evolution, and in this way, we can understand how much the cosmological perturbations deviate from the adiabatic state ($c_{\mathrm{eff}} = c_s$) in different epochs of the Universe.

In the present work, we will further apply the results of the linear regime of perturbations for growth factor to examine the variation of the Integrated Sachs–Wolfe (ISW) \cite{Sachs1967}. We study the consequences of the clustering DBI dark energy model in the ISW effect and compare the results with the $\Lambda$CDM model. Furthermore, we study consequences of our model in spherical collapse scenario and compare the results with $\Lambda$CDM and EdS models.

In the following, we examine the growth of the perturbations in both the linear and non-linear regimes. In our examination, we assume that the dynamics of the DBI scalar field is determined by the self-interacting quartic potential $V(\phi) = \lambda \phi^4 /4$ where $\lambda$ is a constant. We also take the DBI warp factor in the anti-de sitter form $f(\phi)=f_0/\phi^4 $ with constant $f_0$, and study the cosmological implications of our model. For this purpose, we first investigate the consequences of our model in the background cosmology in Sec. \ref{section:background}. Then, in Sec. \ref{section:linear}, we examine the growth of the perturbations at the linear level. Subsequently, in Sec. \ref{section:spherical_collapse}, we proceed to study the non-linear dynamics of DM and DE overdensities by using of SCM. Finally, in Sec. \ref{section:conclusions}, we present our concluding remarks.


\section{DBI dark energy model}
\label{section:background}

The action of DBI dark energy in the context of Einstein gravity is in the form \cite{Silverstein2004}
\begin{equation}
 \label{S}
 S=\int d^{4}x\sqrt{-g}\left[\frac{M_P^2}{2}R+\mathcal{L}(X,\phi)\right]+S_m,
\end{equation}
where $M_{P}=(8\pi G)^{-1/2}$ is the reduced Planck mass. Also $g$ and $R$ are respectively the determinant of the metric $g_{\mu\nu}$ and the Ricci scalar. Furthermore, $S_m$ is the matter field action and $\mathcal{L}(X,\phi)$ is the DBI non-canonical scalar field Lagrangian given by
\begin{equation}
 \label{L}
 \mathcal{L}(X,\phi)\equiv f^{-1}(\phi)\left[1-\sqrt{1-2f(\phi)X}\right]-V(\phi)  .
\end{equation}
In the above relations, $f(\phi)$, $X\equiv - g^{\mu \nu} \partial_\mu \phi~ \partial _\nu \phi/2$ and $V(\phi)$, are respectively the warp factor, the canonical kinetic term, and potential of the DBI scalar field $\phi$.

For the DBI Lagrangian (\ref{L}), the corresponding DE density and pressure are given as follows
\begin{align}
 \label{rhod}
 \rho_d &\equiv 2X\mathcal{L}_{,\rm X} -\mathcal{L}=\frac{\gamma-1}{f(\phi)}+V(\phi),
 \\
 \label{pd}
 p_d &\equiv \mathcal{L}=\frac{\gamma-1}{\gamma f(\phi)}-V(\phi),
\end{align}
where ``$,X\equiv \partial/\partial X$'', and the parameter $\gamma$ is defined as
\begin{equation}
 \label{gamma}
 \gamma\equiv\frac{1}{\sqrt{1-2f(\phi)X}}.
\end{equation}
The $\gamma$ parameter determines the relativistic limit of brane motion in a warped background.

Taking the variation of action (\ref{S}) with respect to the spatially flat Friedmann-Robertson-Walker (FRW) metric gives the first and second Friedmann equations as follows
\begin{align}
 & H^{2}=\frac{1}{3M_{P}^{2}}(\rho_d+\rho_{m})=\frac{1}{3M_{P}^{2}}\left(\frac{\gamma-1}{f(\phi)}+V(\phi)+\rho_{m}\right),
 \label{H}
 \\
 & \dot{H}=-\frac{1}{2M_{P}^{2}}\big(\gamma\dot{\phi}^{2}+\rho_{m}\big),
 \label{Hdot}
\end{align}
where $H=\dot{a}/a$ and $\rho_m$ are respectively the Hubble parameter and pressureless matter energy density. It should be noted that the canonical kinetic term for the flat FRW metric turns into $X=\dot{\phi}^2/2$.

Furthermore, varying the action (\ref{S}) with respect to $\phi$ gives the equation of motion of the scalar field as
\begin{equation}
 \label{phiddot}
 \ddot{\phi}+\frac{3 f_{,\phi}}{2 f}\dot{\phi}^2-\frac{f_{,\phi}}{f^2}+\frac{3H}{\gamma^2}\dot{\phi}+\left(V_{,\phi}+\frac{f_{,\phi}}{f^2}\right)\frac{1}{\gamma^3}=0,
\end{equation}
where ``$,\phi\equiv \partial/\partial \phi$''. Note that one can also extract the equation of motion (\ref{phiddot}) from the continuity equation of the scalar field $\phi$,
\begin{equation}
 \label{rhoddot}
 \dot{\rho}_d+3H(\rho_d+p_d)=0.
\end{equation}
Additionally, from the continuity equation of pressureless DM,
\begin{equation}
 \label{rhomdot}
 \dot{\rho}_m+3H\rho_m=0,
\end{equation}
one can obtain the evolution of matter field,
\begin{equation}
 \label{rhom}
 \rho_m=\rho_{m_0}a^{-3},
\end{equation}
where $\rho_{m_0}$ denotes the present matter energy density at the scale factor $a_0=1$.

In the context of non-canonical scalar field, the propagation speed of the scalar field fluctuations $\delta\phi$ among the homogeneous background is specified by the adiabatic sound speed
\begin{equation}
 \label{cs}
 c^2_s \equiv \frac{p_{d,{\rm X}}}{\rho_{d,{\rm X}}}.
\end{equation}
Note that $c_s$ should satisfy the condition $0<c_{s}^{2}\leq1$ because the adiabatic sound speed must be real and subluminal \cite{Adams2006, Babichev2008, Franche2010}. In DBI dark energy model, one can obtain the sound speed from Eqs. (\ref{rhod}), (\ref{pd}), and (\ref{cs}) as
\begin{equation}
\label{csDBI}
c_{s}=\sqrt{1-2f(\phi)X}=\frac{1}{\gamma}.
\end{equation}

It should be noted that only three equations from Eqs. (\ref{H}), (\ref{Hdot}), (\ref{phiddot}), and (\ref{rhomdot}) are independent. Throughout this paper, we regard the set of Eqs. (\ref{H}), (\ref{Hdot}), and (\ref{rhomdot}) as the independent equations describing the dynamics of the Universe thoroughly. 

To specify the potential governing the dynamics of the DBI scalar field in our dark energy scenario, we note that it is essential that the action (\ref{S}) to be invariant under some group of internal symmetries. In particular, in order to the action will remain invariant under the transformation $\phi \to -\phi$, the potential $V(\phi)$ should be an even function. A conventional choice satisfying this condition has the following form 
\begin{equation}
 \label{VV0m}
 V(\phi)=V_{0}-\frac{1}{2}m^{2}\phi^{2}+\frac{1}{4}\lambda\phi^{4},
\end{equation}
where $V_0$, $m$, and $\lambda$ are constant parameters. The parameter $m$ represents mass of the scalar field, while $\lambda$ is known as the self-interacting coupling. Furthermore, it should be noted that the symmetry is essentially broken spontaneously at an energy scale which is quite higher than the energy scale of the present Universe, and this fact has a special importance in different aspects of theoretical cosmology \cite{Liddle2000}. If we take the potential parameters in potential (\ref{VV0m}) as $V_{0}=\lambda\mu^{4}/4$ and $m^{2}=\lambda\mu^{2}$, then it turns into
\begin{equation}
 \label{Vmu}
 V(\phi)=\frac{1}{4}\lambda\left(\phi^{2}-\mu^{2}\right)^{2}.
\end{equation}
This potential has a Mexican-hat shape which can explain the process of spontaneous symmetry breaking satisfactorily \cite{Liddle2000}. The minima of the above potential which lie at $\phi=\pm\mu$, represent the possible vacuum values of the scalar field, which are called vacuum expectation values in the quantum field theory. Since oscillations of the field is not symmetric around the minima anymore, it is said that the symmetry is spontaneously broken. The parameters $\mu$ and $V_0$ are related to the mass $m$ of the scalar field via $\mu=m/\sqrt{\lambda}$ and $V_{0}=m^{4}/4\lambda$, respectively. Therefore, if we suppose that the vacuum expectation value $\mu$ is much less than the value of $\phi$, then the potential (\ref{Vmu}) can be approximated by the quartic potential
\begin{equation}
 \label{V}
 V(\phi) \approx \frac{1}{4} \lambda \phi^{4}.
\end{equation}
In this paper, we work with values of the scalar field which are larger than $M_P$, and also we deal with values of self-interacting coupling parameter as $\lambda\gtrsim\mathcal{O}(1)$. If the mass $m$ of the scalar field is supposed to be much less than $M_P$, then the condition $\mu=m/\sqrt{\lambda}\ll\phi$ is satisfied perfectly, and consequently, the potential can be approximated consistently by the quartic form given in Eq. (\ref{V}).

Another point that motives us to consider the quartic potential (\ref{V}) in our investigation, is the fact that in our previous paper \cite{Fahimi2018}, we noticed that with constant values of $c_s$ much less than the light speed, the DBI potential behaves asymptotically as $V(\phi)\propto\phi^{4}$. In the present work, although $c_s$ varies with time, but its value remains much less than the light speed along the whole of the cosmic evolution. So it makes sense to take the quartic form (\ref{V}) for the potential up to a good approximation.

We also assume that the DBI warp factor follows the anti-de Sitter (AdS) form \cite{Aharony2000, Silverstein2004, Alishahiha2004}
\begin{equation}
 \label{f}
 f(\phi)=\frac{f_{0}}{\phi^{4}},
\end{equation}
with another positive constant $f_0$. Therefore, by applying Eqs. (\ref{V}) and (\ref{f}) in Eqs. (\ref{gamma}), (\ref{H}), and (\ref{Hdot}), the background equations turn into
\begin{align}
 \gamma &=f_{0}\phi^{-4}\left(3M_{P}^{2}H^{2}-\rho_{m0}a^{-3}\right)-\frac{\lambda f_{0}}{4}+1,
 \label{gamma-phi}
 \\
 2M_{P}^{2}aHH' &=-\rho_{m0}a^{-3}-\frac{\left(\gamma^{2}-1\right)\phi^{4}}{f_{0}\gamma},
 \label{Hprime}
 \\
 aH\phi' &=-\frac{\sqrt{\gamma^{2}-1}\phi^{2}}{\sqrt{f_{0}}\gamma},
 \label{phiprime}
\end{align}
where the prime denotes the derivative with respect to the scale factor $a$. Here, it is useful to introduce the following dimensionless quantities
\begin{equation}
 \label{dimensionless-quantities}
 \tilde{H}\equiv\frac{H}{H_{0}}  , \qquad \tilde{\phi}\equiv\frac{\phi}{M_{P}} , \qquad \tilde{\lambda}\equiv\frac{H_{0}^{2}\lambda}{M_{P}^{2}}  , \qquad \tilde{f_{0}}\equiv\frac{H_{0}^{2}f_{0}}{M_{P}^{2}}  .
\end{equation}
As a result, Eqs. (\ref{gamma-phi}), (\ref{Hprime}), and (\ref{phiprime}) can be rewritten in the dimensionless form as
\begin{align}
 \gamma &=3\tilde{f}_{0}\tilde{\phi}^{-4}\left(\tilde{H}^{2}-\Omega_{m_{0}}a^{-3}\right)-\frac{\tilde{\lambda}\tilde{f}_{0}}{4}+1  ,
 \label{gamma-phitilde}
 \\
 2\tilde{H}\tilde{H}^{\prime} &=-3\Omega_{m0}a^{-4}-\frac{\left(\gamma^{2}-1\right)\tilde{\phi}^{4}}{\tilde{f}_{0}\gamma}  ,
 \label{Htildeprime}
 \\
 a\tilde{H}\tilde{\phi}^{\prime} &=-\frac{\sqrt{\gamma^{2}-1}\tilde{\phi}^{2}}{\sqrt{\widetilde{f}_{0}}\gamma}  .
 \label{phitildeprime}
\end{align}

\begin{figure}
 \centering
 \includegraphics[scale=0.4]{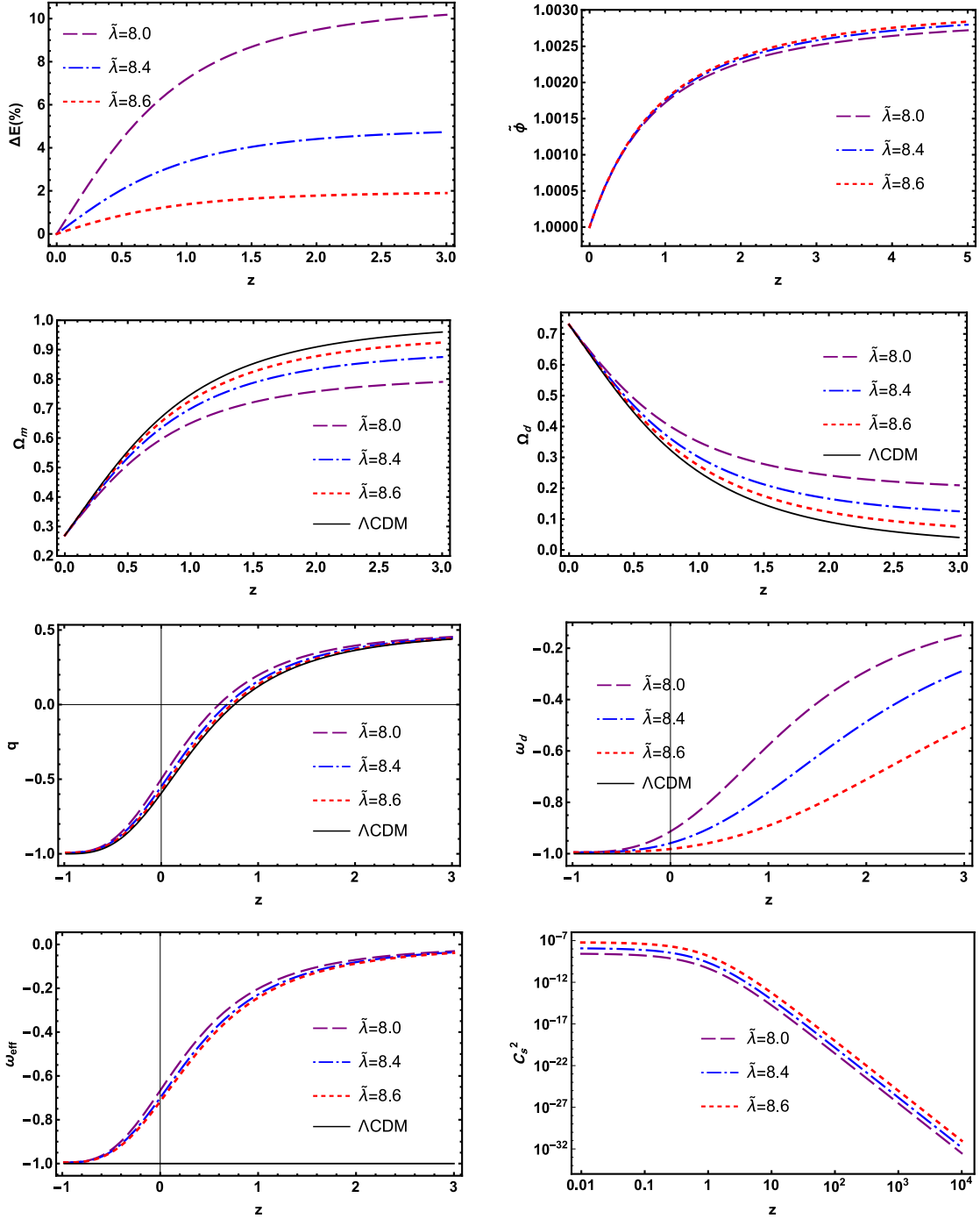}
 \caption{\small{Variations of the relative deviation $ \Delta E(z)$ of the normalized Hubble parameter for the DBI models in comparison with the $\Lambda$CDM, the normalized DBI scalar field $\tilde{\phi}$, the DM density parameter $\Omega_m$, the DE density parameter $\Omega_d$, the deceleration parameter $q$, the EoS parameter of DE $\omega_d$, the effective EoS parameter $\omega_{\rm eff}$ and the adiabatic sound speed $c^2_s$. Auxiliary parameters are $\Omega_{m_0}=0.27$ and $\tilde{f}_0=10^5$.}}
 \label{cosmofig1}
\end{figure}

\begin{figure}
\centering
\includegraphics[scale=0.4]{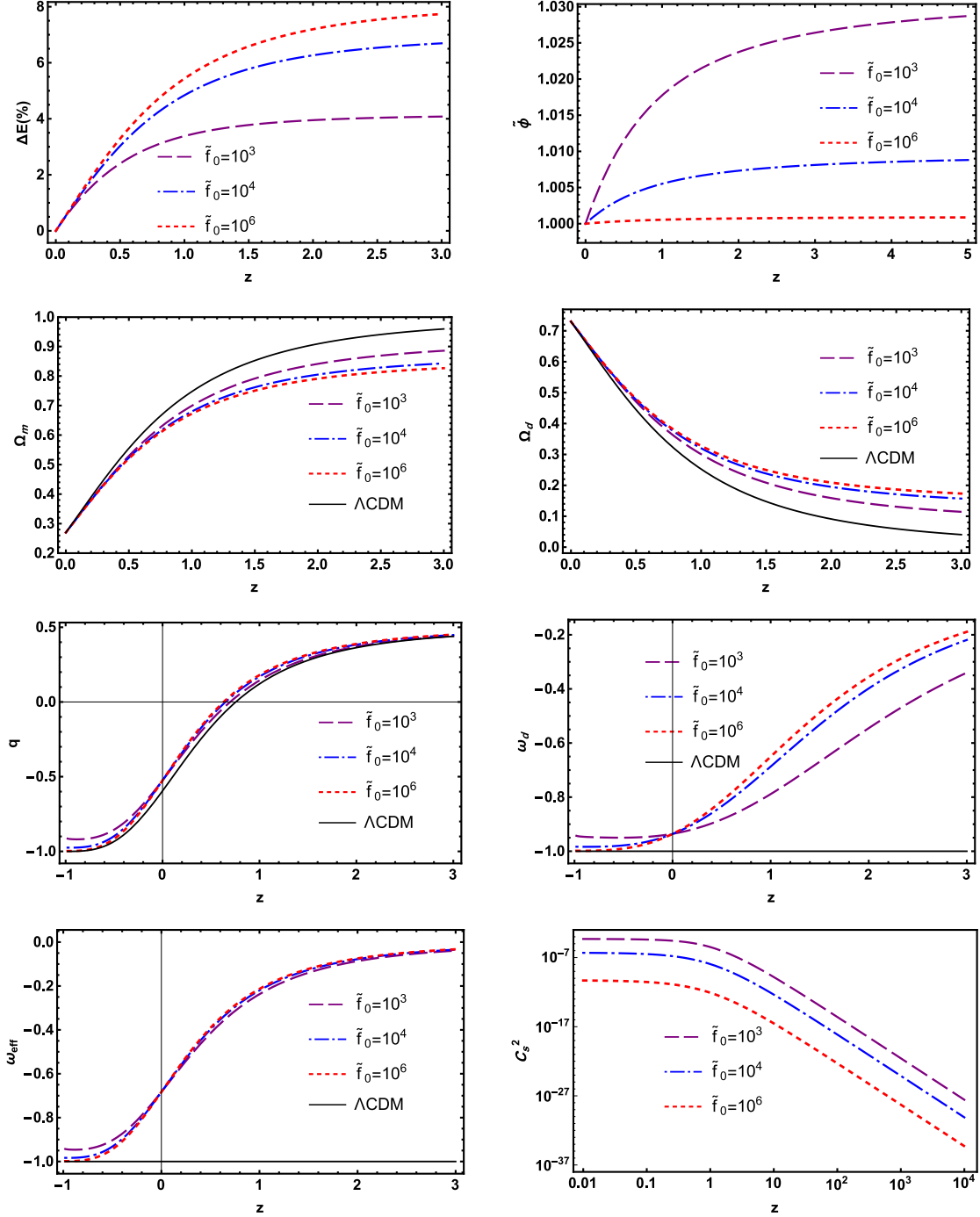}
\caption{\small{Same as Fig. \ref{cosmofig1}, but for $\tilde{\lambda}=8.2$ and different values of $\tilde{f}_0$.}}
\label{cosmofig2}
\end{figure}

Now, we solve Eqs. (\ref{Htildeprime}) and (\ref{phitildeprime}) numerically for some typical values of $\tilde{\lambda}$ and $\tilde{f}_0$ with the initial conditions $\tilde{\phi}(a_0)=1$ and $\tilde{H}(a_0)=1$. Also, we set $a_0=1$ and $\Omega_{m_0}=0.27$. The plots of the evolutionary behaviours of $\Delta E \equiv \Delta \tilde{H}=100\left(\frac{E_{{\rm DBI}}}{E_{\Lambda{\rm CDM}}}-1\right)$ and $\tilde{\phi}$ are shown versus redshift $z=\frac{1}{a}-1$ in Fig. \ref{cosmofig1} for $\tilde{\lambda}=(8,8.4,8.6)$ and $\tilde{f}_0=10^5$, and in Fig. \ref{cosmofig2} for $\tilde{f}_0=(10^3,10^4,10^6)$ and $\tilde{\lambda}=8.2$. The figures show that (i) the Hubble parameter in our model is larger than the result of the $\Lambda$CDM model, since $\Delta E >0$. We see that $\Delta E$ is more pronounced for smaller $\tilde{\lambda}$ and larger $\tilde{f}_0$. (ii) The normalized DBI scalar field $\tilde{\phi}$ descends when we go towards the low redshifts.

Besides, using the numerical results of $\tilde{H}$ and $\tilde{\phi}$, one can obtain the evolution of other background quantities including the dimensionless DM and DE density parameters ($\Omega_m$, $\Omega_d$), the deceleration parameter $q=-1-\dot{H}/H^2$, the equation of state (EoS) parameter of DE $\omega_d\equiv p_d/\rho_d$, the effective EoS parameter $\omega_{\rm eff}=-1-\frac{2}{3}\frac{\dot{H}}{H^2}$, and the adiabatic sound speed $c_s$ . The results are illustrated in Figs. \ref{cosmofig1} and \ref{cosmofig2}. The figures present that (i) $\Omega_d$ and $\Omega_m$ increase and decrease, respectively, with decreasing the redshift; (ii) The deceleration parameter $q$ for all the models except $\tilde{f}_0=10^3$, starts from an early matter-dominated regime $(q = 0.5)$ and it behaves like the de Sitter universe ($q = -1$) in the late time ($z \rightarrow -1$), as expected. Also, the deceleration parameter in the near past illustrates a transition from a decelerating phase ($q > 0$) to a cosmic acceleration ($q < 0$). Here, the transition redshift $z_t$ is smaller than the one in the $\Lambda$CDM model, $z_t^{\Lambda \rm CDM}=0.755$. In our DBI model, we obtain the transition redshifts $z_t = (0.593,0.678,0.724$) for $\tilde{\lambda}= (8,8.4,8.6)$ with $\tilde{f}_0=10^5$ (see Fig. \ref{cosmofig1}) and $z_t=(0.697,0.648,0.630)$ for $\tilde{f}_0=(10^3,10^4,10^6)$ with $\tilde{\lambda}=8.2$ (see Fig. \ref{cosmofig2}); (iii) The EoS parameter $\omega_d$ of our DBI model behaves like that of the quintessence DE ($\omega_d>-1$) during history of the Universe. This point is in agreement with that obtained by \cite{Devi2011} for the tachyon DE model; (iv) The effective EoS parameter, $\omega_{\rm eff}$, varies from an early matter-dominant epoch ($\omega_{\rm eff}=0$) to the $\Lambda$CDM regime ($\omega_{\rm eff} \rightarrow -1$) in the late-time future. We see that in the case $\tilde{f}_0=10^3$ the model does not behave like $\Lambda$CDM in the late time; (v) The adiabatic sound speed evolves with time, and for the all models it remains very small during the cosmological evolution. We see that with larger values of $\tilde{f}_0$, it goes toward the smaller values.

\section{Linear perturbation theory}
\label{section:linear}

In this section, we study the linear regime of structure formation in the DBI dark energy scenario. For this purpose, it is suitable to start with the perturbations of the DBI scalar field. Our analysis is concise here, and for more details on the perturbations of scalar fields with non-canonical Lagrangian, one can see \cite{Garriga1999, Abramo2004, Amendola2004, Tsujikawa2005, Bertacca2007, Mukhanov2005}. Perturbing the background metric leads to a perturbation for the scalar field which is denoted here by $\varphi\equiv\delta\phi$. Like the scalar field $\phi$, we normalize its perturbation to $M_P$, so $\tilde{\varphi}\equiv\varphi/M_{P}$. Due to perturbing the scalar field $\phi$, its energy density and pressure is also perturbed as
\begin{align}
 \label{deltarhodXphi}
 \delta\rho_{d} & =\frac{\partial\rho_{d}}{\partial X}\delta X+\frac{\partial\rho_{d}}{\partial\phi}\varphi,
 \\
 \label{deltapdXphi}
 \delta p_{d} & =\frac{\partial p_{d}}{\partial X}\delta X+\frac{\partial p_{d}}{\partial\phi}\varphi.
\end{align}
For $\delta X$ in the above equations, we can use \cite{Mukhanov2005, Amendola2010}
\begin{equation}
 \label{deltaX}
 \delta X=a^{2}H^{2}\phi'\left(\varphi'-\phi'\Phi\right),
\end{equation}
where $\Phi$ is the gravitational potential appeared due to the metric perturbation. We remind that the prime denotes the derivative with respect to scale factor $a$. Using Eqs. (\ref{rhod}), (\ref{pd}), and (\ref{deltaX}) in Eqs. (\ref{deltarhodXphi}) and (\ref{deltapdXphi}), the perturbed energy density and pressure of the DBI scalar field are resulted in as
\begin{align}
 \delta\rho_{d}= & \frac{a^{2}H^{2}\phi'\left(\varphi'-\phi'\Phi\right)}{(1-2fX)^{3/2}}+\varphi\left[\frac{f_{,\phi}}{f^{2}}\left(1-\frac{1}{\sqrt{1-2fX}}\right)+\frac{f_{,\phi}X}{f(1-2fX)^{3/2}}+V_{,\phi}\right],
 \label{deltarhophi}
 \\
 \delta p_{d}= & -\frac{a^{2}H^{2}\phi'\left(\varphi'-\phi'\Phi\right)f^{2}}{\sqrt{1-2fX}}
 \nonumber
 \\
 & -\varphi\left[\frac{f_{,\phi}}{f^{2}(1-2fX)}\left(\left[fX\left(\sqrt{1-2fX}-2\right)-\sqrt{1-2fX}+1\right]\right)+V_{,\phi}\right].
 \label{deltaPphi}
\end{align}
From the above equations, we can show that the effective sound speed $c_{\mathrm{eff}}^{2}\equiv\delta p_{d}/\delta\rho_{d}$ of the DBI model in the presence of the scalar field perturbations is given by
\begin{align}
 c_{\mathrm{eff}}^{2}= & \left(1-2fX\right)\bigg\{ a^{2}H^{2}f^{2}\phi'\left(\Phi\phi'-\varphi'\right)
 \nonumber
 \\
 & +\varphi\left[f_{,\phi}\left(\sqrt{1-2fX}+fX-1\right)+f^{2}V_{,\phi}\sqrt{1-2fX}\right]\bigg\}\bigg\{ a^{2}H{}^{2}f^{2}\phi'\left(\Phi\phi'-\varphi'\right)
 \nonumber
 \\
 & +\varphi\left(f_{,\phi}\left[fX\left(2\sqrt{1-2fX}-3\right)-\sqrt{1-2fX}+1\right]-f^{2}V_{,\phi}(1-2fX)^{3/2}\right)\bigg\}^{-1}.
 \label{ceff}
\end{align}
Since the effective sound speed is related to the perturbed quantities, thus it will be a gauge-dependent quantity. In different gauges, the velocity divergence
\begin{equation}
 \label{theta}
 \theta\equiv i\mathbf{k.v},
\end{equation}
takes different values. Instead of $\theta$ is more appropriate to work with its normalized form,
\begin{equation}
 \label{thetatilde}
 \tilde{\theta}\equiv\frac{\theta}{\mathcal{H}}=\frac{\theta}{aH}.
\end{equation}
The normalized velocity divergence of the scalar field has relation with its perturbation as follows \cite{Mukhanov2005, Amendola2010}
\begin{equation}
 \label{varphithetatilded}
 \tilde{\varphi}=\frac{a^{3}H^{2}}{k^{2}}\tilde{\phi}'\tilde{\theta}_{d}.
\end{equation}
In the rest frame of the scalar field (i.e. in the gauge where the scalar field is at rest), we have $\tilde{\theta}_{d}=0$, and so from the above equation, we will have $\tilde{\varphi}=0$. Inserting this into Eq. (\ref{ceff}), we see
\begin{equation}
 \label{ceffcs}
 c_{\mathrm{eff}}^{2}=c_{s}^{2}=1-2fX.
\end{equation}
Therefore, in the rest frame of DE, the effective sound speed (\ref{ceff}) of the DBI scalar field reduces to the adiabatic sound speed which is already given in Eq. (\ref{cs}).

We derived the full expression of the effective sound speed for the DBI dark energy model in Eq. \eqref{ceff}. We will apply this expression in the equations governing the perturbations evolution, and hence our work is preferred relative to most of the previous works which does not calculate this quantity explicitly, and consider exclusively the two cases of non-clustering ($c_{\mathrm{eff}}=1$) and full-clustering ($c_{\mathrm{eff}}=0$) DE scenarios. The approach we followed here to extract $c_{\mathrm{eff}}$ for our model is also applicable for the other $k$-essence DE models. Having the relation of $c_{\mathrm{eff}}$ at hand, we can also examine the evolution of the effective sound speed during the cosmological eras, and specially compare it with the adiabatic sound speed in different epochs. In this way, we can specify the extent that the cosmological perturbations deviates from the adiabatic state as the Universe evolves.

In order to investigate the linear growth rate of density perturbations of both the pressureless DM and DBI scalar field in our model, we use the Pseudo-Newtonian (PN) formalism \cite{Hwang1997, Lima1997, Hwang2006}. In this formalism, the linear density contrasts of non-relativistic DM $\delta_m\equiv \delta\rho_m/\rho_m$ and DE $\delta_d\equiv \delta\rho_\phi/\rho_\phi$ satisfy the following equations \cite{Abramo2009}
\begin{align}
 \label{dotdeltam}
 & \dot{\delta}_m+\frac{\theta_m}{a}=0,
 \\
 \label{dotdeltad}
 & \dot{\delta}_d+(1+\omega_d)\frac{\theta_d}{a}+3H\big(c_{\rm eff}^2-\omega_d\big)\delta_d =0,
 \\
 \label{thetam}
 & \dot{\theta}_{m}+H\theta_{m}-\frac{k^{2}}{a}\Phi=0,
 \\
 \label{thetad}
 & \dot{\theta}_{d}+H\theta_{d}-\frac{k^{2}c_{{\rm eff}}^{2}\delta_{d}}{a(1+\omega_{d})}-\frac{k^{2}}{a}\Phi=0.
\end{align}
For $\Phi$ in the above equation, we use the Poisson equation which on the sub-horizon scales and in the Fourier space reads \cite{Lima1997}
\begin{equation}
 \label{Phi}
 -\frac{k^{2}}{a^{2}}\Phi=\frac{3}{2}H^{2}\big[\Omega_{m}\delta_{m}+\big(1+3~c_{{\rm eff}}^{2}\big)\Omega_{d}\delta_{d}\big].
\end{equation}
To determine the evolutionary behaviours of DM and DE perturbations, it is more convenient to express Eqs. (\ref{dotdeltam})-(\ref{thetad}) in terms of the scale factor $a$. In this regards, we get
\begin{align}
 \label{dadeltam}
 & \delta_{m}^{\prime}+\frac{\tilde{\theta}_{m}}{a}=0,
 \\
 \label{dadeltad}
 & \delta_{d}^{\prime}+\frac{3}{a}\big(c_{{\rm eff}}^{2}-\omega_{d}\big)\delta_{d}+(1+\omega_{d})\frac{\tilde{\theta}_{d}}{a}=0,
 \\
 \label{lithetam}
 & \tilde{\theta}_{m}^{\prime}+\left(\frac{2}{a}+\frac{H^{\prime}}{H}\right)\tilde{\theta}_{m}+\frac{3}{2a}\left[\Omega_{m}\delta_{m}+\big(1+3c_{{\rm eff}}^{2}\big)\Omega_{d}\delta_{d}\right]=0,
 \\
 \label{lithetad}
 & \tilde{\theta}_{d}^{\prime}+\left(\frac{2}{a}+\frac{H^{\prime}}{H}\right)\tilde{\theta}_{d}-\frac{k^{2}c_{{\rm eff}}^{2}\delta_{d}}{a^{3}H^{2}(1+\omega_{d})}+\frac{3}{2a}\left[\Omega_{m}\delta_{m}+\big(1+3c_{{\rm eff}}^{2}\big)\Omega_{d}\delta_{d}\right]=0.
\end{align}
In order to solve this set of coupled differential equations, we assume that at the initial redshift $z_i=10^4$, the following initial conditions are valid in the Einstein-de Sitter (EdS) limit \cite{Batista2013, Pace2014}
\begin{align}
 \label{deltami}
 & \delta_{m_i} = a_i=(1+z_i)^{-1},
 \\
 \label{deltadi}
 & \delta_{d_i}=\left(\frac{1+\omega_{d_{i}}}{1-3\omega_{d_{i}}}\right)\delta_{m_{i}},
 \\
 \label{thetami}
 & \tilde{\theta}_{m_{i}}=-\delta_{m_{i}}.
\end{align}
To derive Eq. (\ref{deltami}), we note that during the matter-dominated era, the density contrast of DM evolve as $\delta_m = C a$, where the proportional constant can bet set $C=1$ without any loss of generality. Equations (\ref{deltadi}) and (\ref{thetami}) can simply be derived respectively from Eqs. (\ref{dadeltam}) and (\ref{lithetam}) in the matter-dominated era. We also can extract an initial condition for $\tilde{\theta}_{d_{i}}$ by using Eq. (\ref{dadeltad}). To this aim, we assume that at the initial times the scalar field perturbation $\tilde{\varphi}=0$ is so small that the effective sound speed (\ref{ceff}) is approximately equal to the adiabatic sound speed ($c_{{\rm eff}}\approx c_{s}$).

\begin{figure}
\begin{minipage}[b]{1\textwidth}
\subfigure[\label{Da1} ]{ \includegraphics[width=.48\textwidth]%
{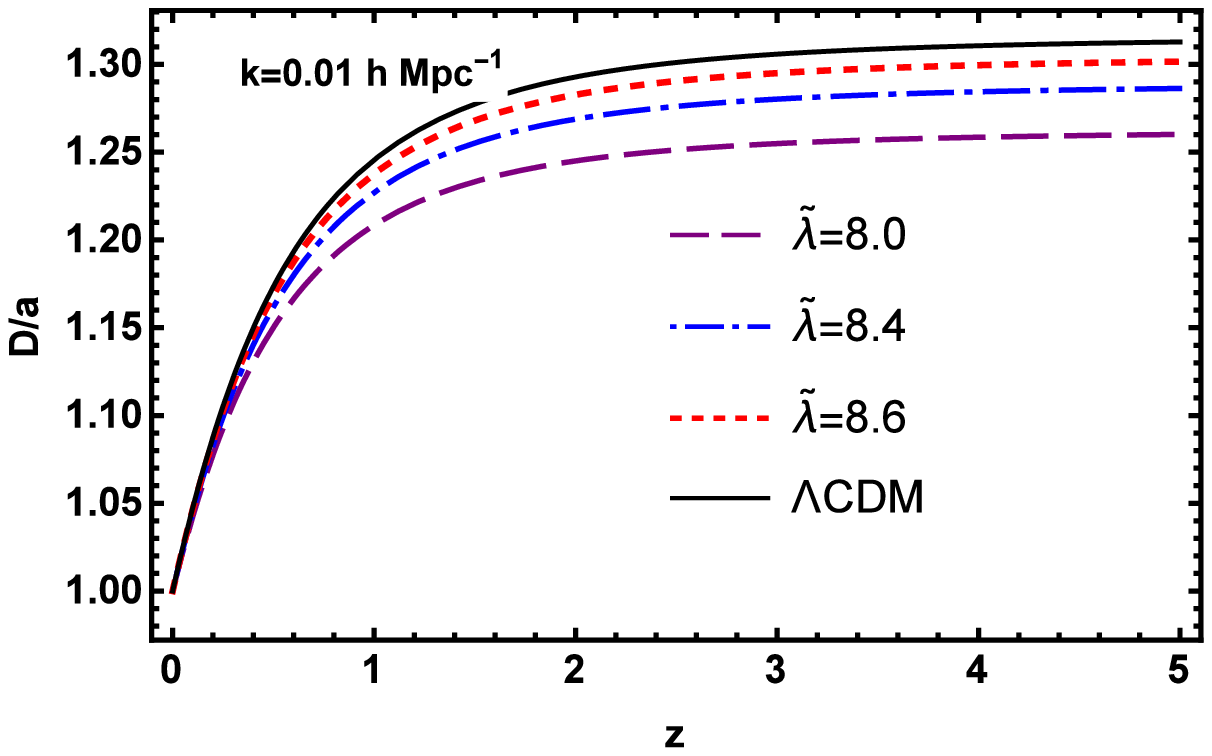}}\hspace{.1cm}
\subfigure[\label{Da2}]{ \includegraphics[width=.48\textwidth]%
{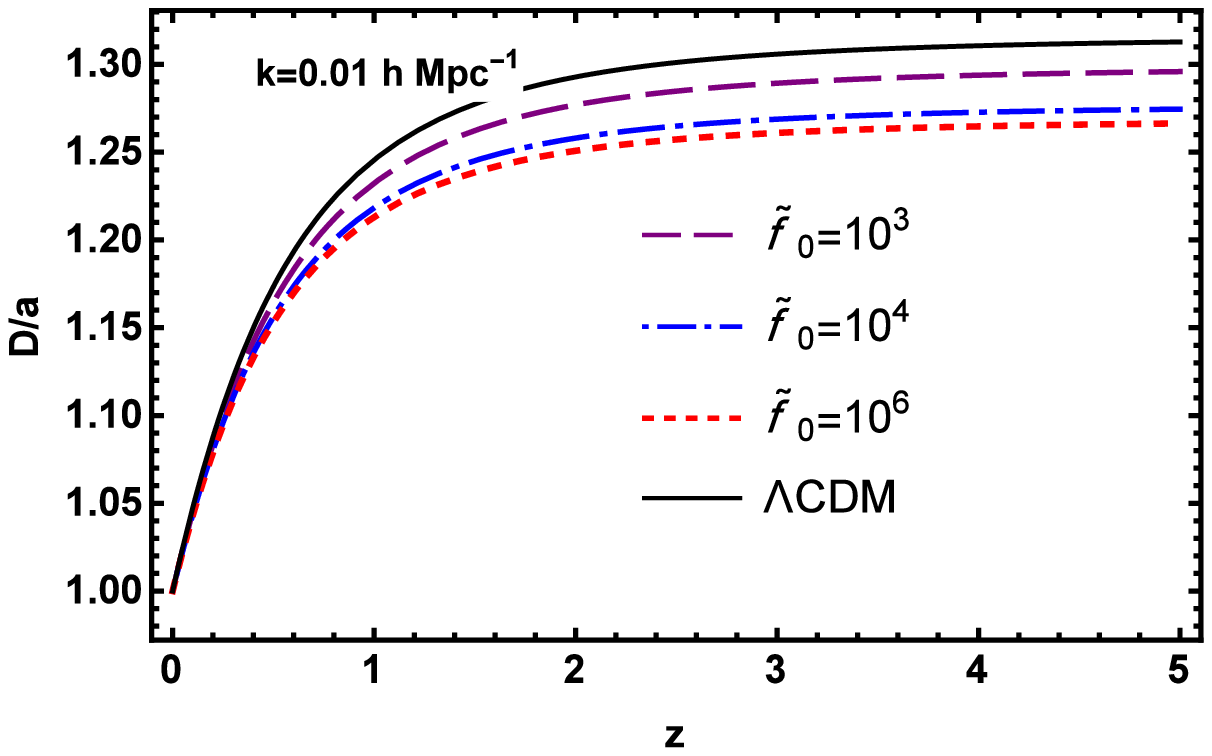}}
\subfigure[\label{Da3}]{ \includegraphics[width=.48\textwidth]%
{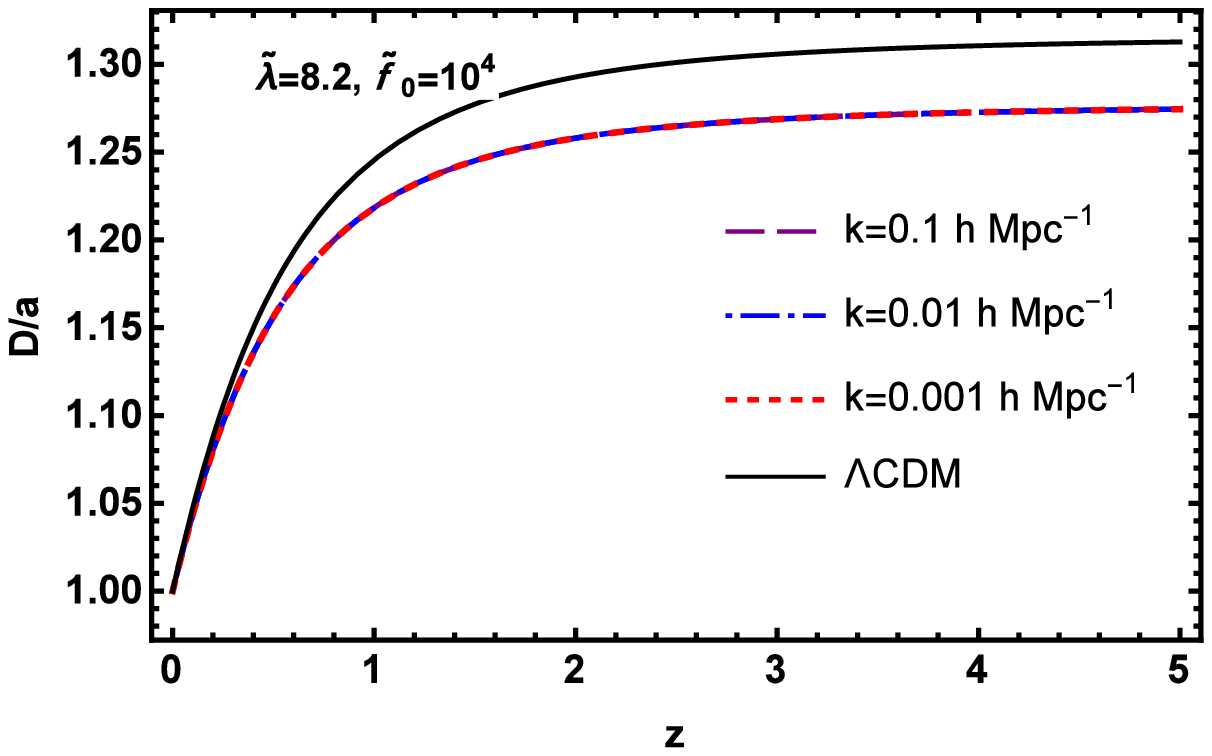}}
\end{minipage}
\caption{(a) Evolution of the growth function relative to its value in a pure
matter model $D/a$, in which $D=\delta_m/\delta_{m_0}$. Auxiliary parameters are $\Omega_{m_0}=0.27$ and $\tilde{f}_0=10^5$. (b) Same as Fig. \ref{Da1}, but for $\tilde{\lambda}=8.2$. (c) Same as Fig. \ref{Da1}, but for $\tilde{\lambda}=8.2$, $\tilde{f}_0=10^5$ and different $k$.
  }\label{Da}
\end{figure}

In Fig. \ref{Da}, we represent the evolution of the growth factor $D=\delta_m/\delta_{m_0}$ normalized to its value in a pure matter model ($D=a$) versus redshift $z$. From the figures we infer that (i) for a given redshift $z$, the result of $D/a$ in the DBI models is smaller than $\Lambda$CDM model; (ii) In the DBI models, like the $\Lambda$CDM model, the growth of perturbations is suppressed at low redshifts, due to the effects of DE which its role becomes dominant around the present redshift; (iii) At large redshifts, $D/a$ tend to a constant value, since in the early periods the influence of DE can be neglected in all the models. It is notice to mention that from Eqs. (\ref{dadeltam})-(\ref{thetami}) although the case of effective sound speed introduces a scale-dependence problem, Fig. \ref{Da3} shows that $D/a$ is almost scale independent. This is because of that the term proportional to $k^2c_{\rm eff}^2$ appeared in Eq. (\ref{lithetad}) is negligible for the small values of $c_{\rm eff}^2$.

\begin{figure}
\begin{minipage}[b]{1\textwidth}
\subfigure[\label{phia} ]{ \includegraphics[width=.48\textwidth]%
{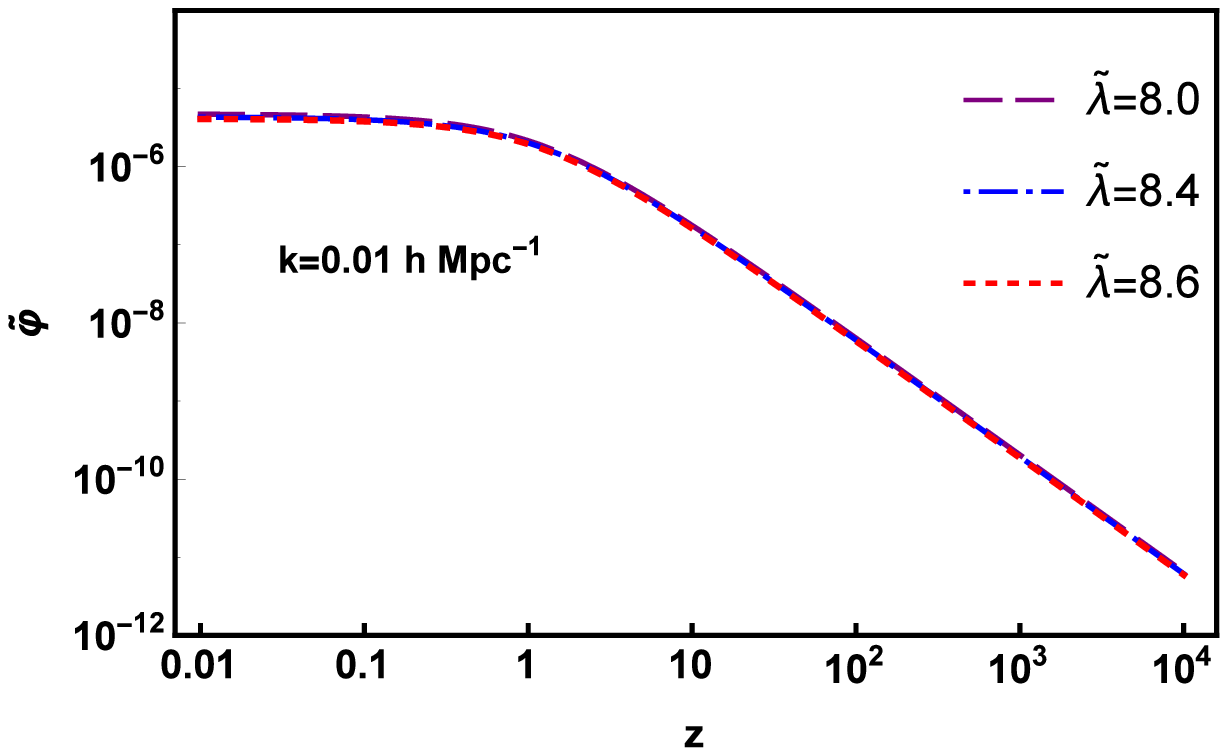}}\hspace{.1cm}
\subfigure[\label{Ceffa}]{ \includegraphics[width=.48\textwidth]%
{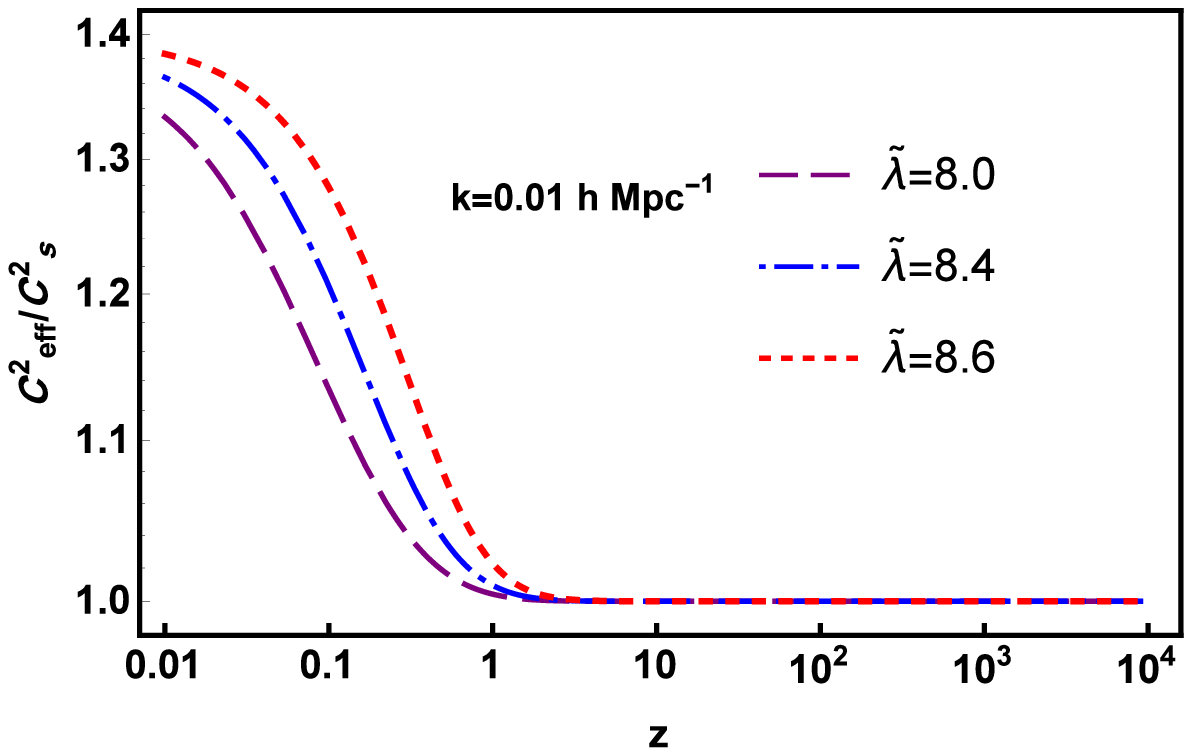}}
\subfigure[\label{phib} ]{ \includegraphics[width=.48\textwidth]%
{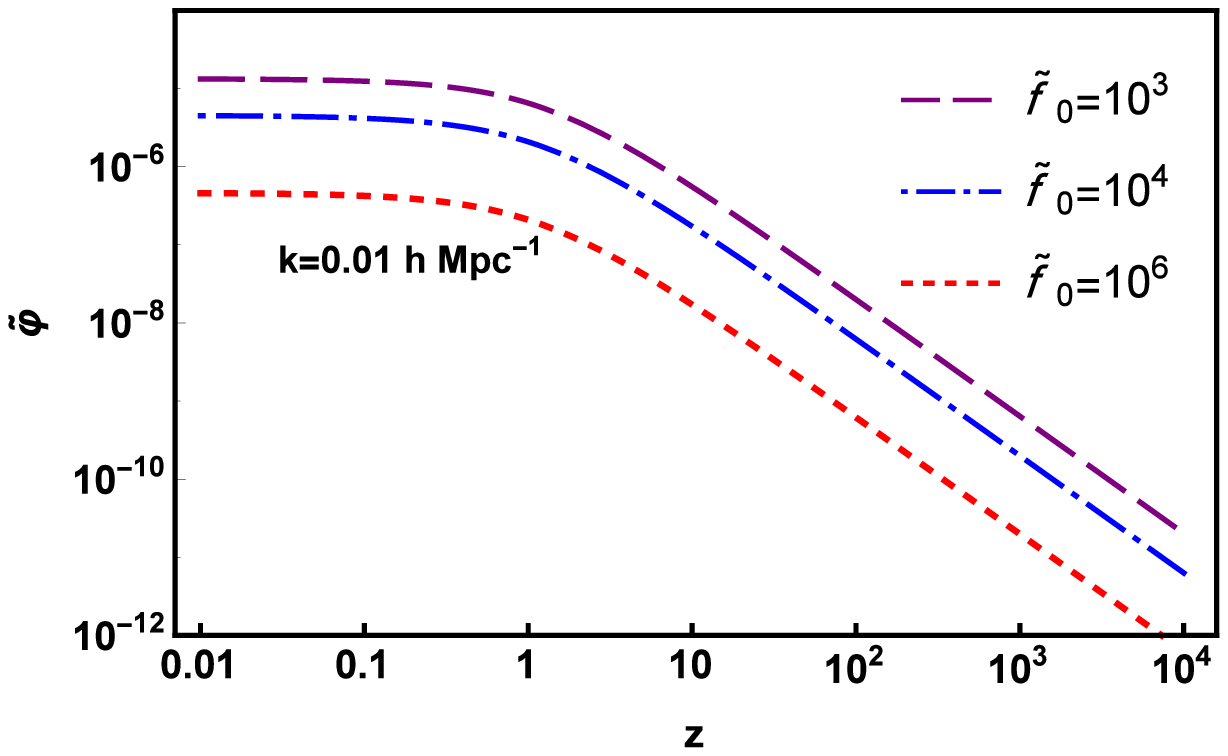}}\hspace{.1cm}
\subfigure[\label{Ceffb}]{ \includegraphics[width=.48\textwidth]%
{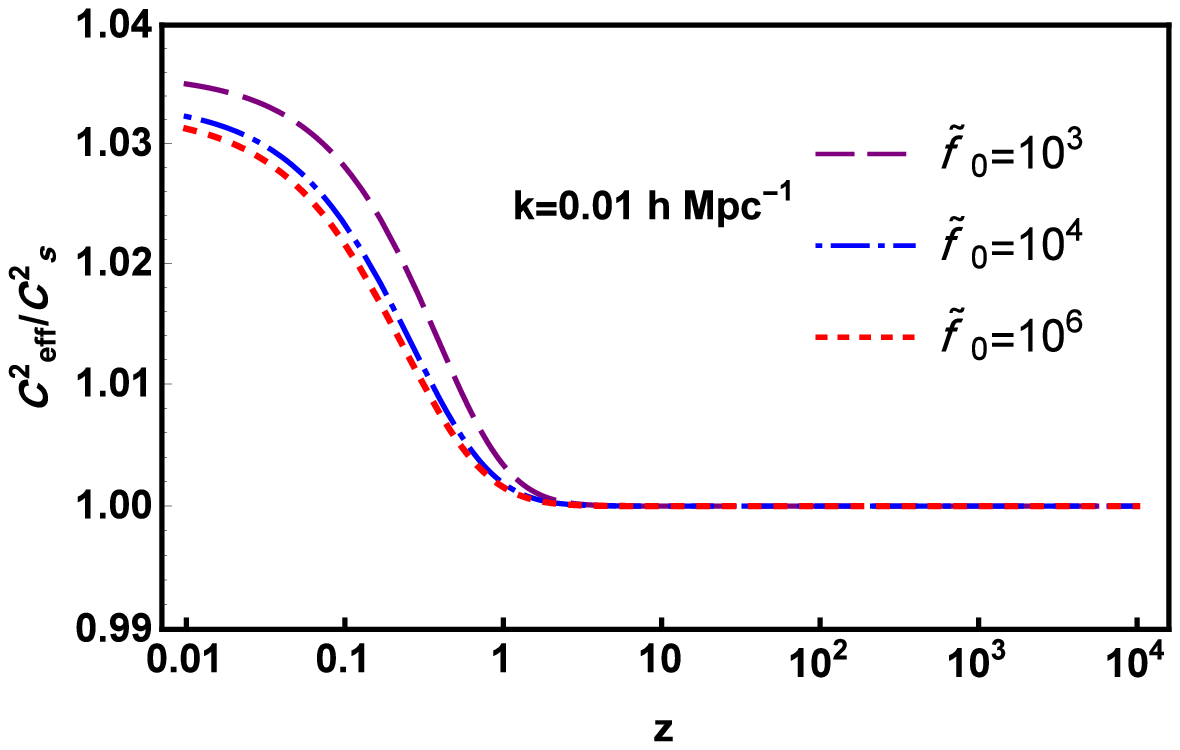}}
\end{minipage}
\caption{(a) Evolution of the normalized scalar field perturbation and (b) the effective sound speed relative to adiabatic one as a function of redshift for different $\tilde{\lambda}$, (c) same as Fig \ref{phia}, but for different $\tilde{f}_0$, (d) same as Fig \ref{Ceffa}, but for different $\tilde{f}_0$. Auxiliary parameters are $k=0.01~\rm{h~Mps^{-1}}$ and $\Omega_{m_0} =0.27$.}
\label{figure:varphitilde}
\end{figure}

In Fig. \ref{phia} and \ref{phib}, the behavior of the normalized scalar field perturbation, $\tilde{\varphi}$, is shown versus redshift. In this figure, it is apparent that the scalar field perturbation begins from small values at the initial times, and then it grows slightly, so that its value remains small until the present epoch. Thus, from Eq. (\ref{ceffcs}), we expect that the effective sound speed (\ref{ceff}) be almost equal to the adiabatic sound speed (\ref{cs}). This fact is illustrated in Fig. \ref{Ceffa} and \ref{Ceffb}. This is an important result implying that in our model, the DE perturbations remains almost adiabatic during history of the Universe.

\begin{figure}
\begin{minipage}[b]{1\textwidth}
\subfigure[\label{phic} ]{ \includegraphics[width=.48\textwidth]%
{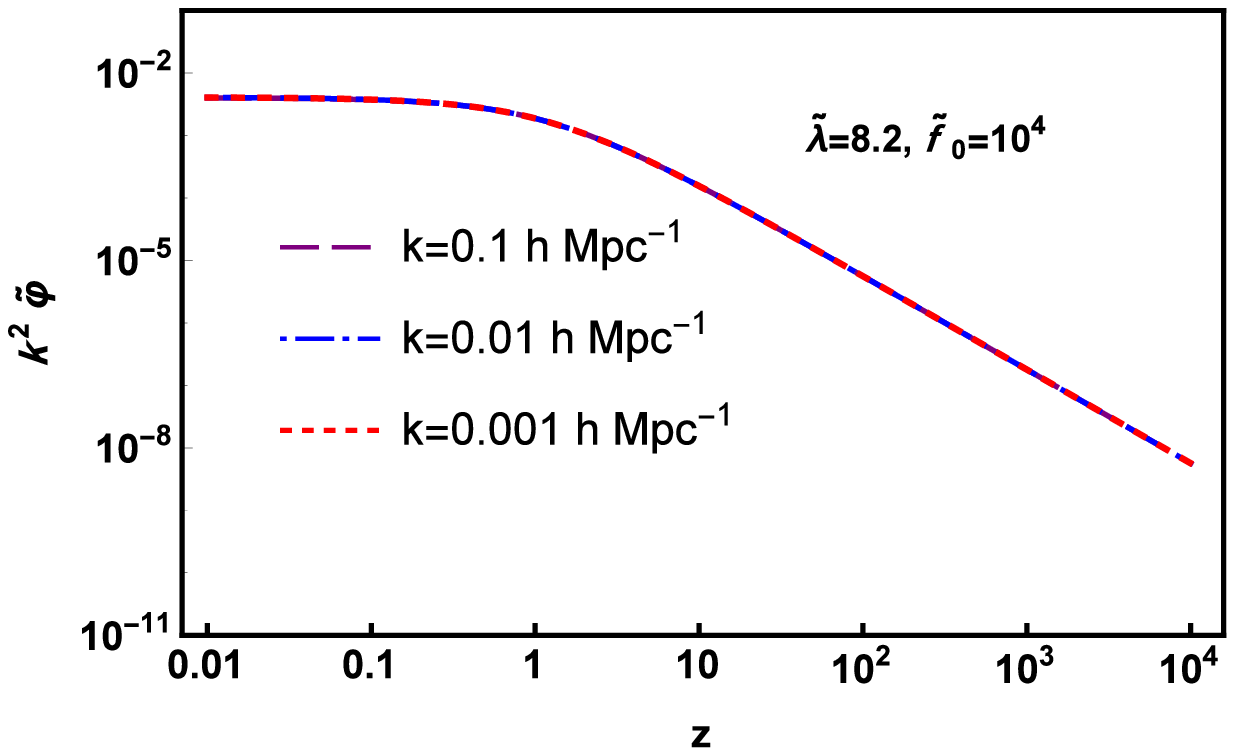}}\hspace{.1cm}
\subfigure[\label{Ceffc}]{ \includegraphics[width=.48\textwidth]%
{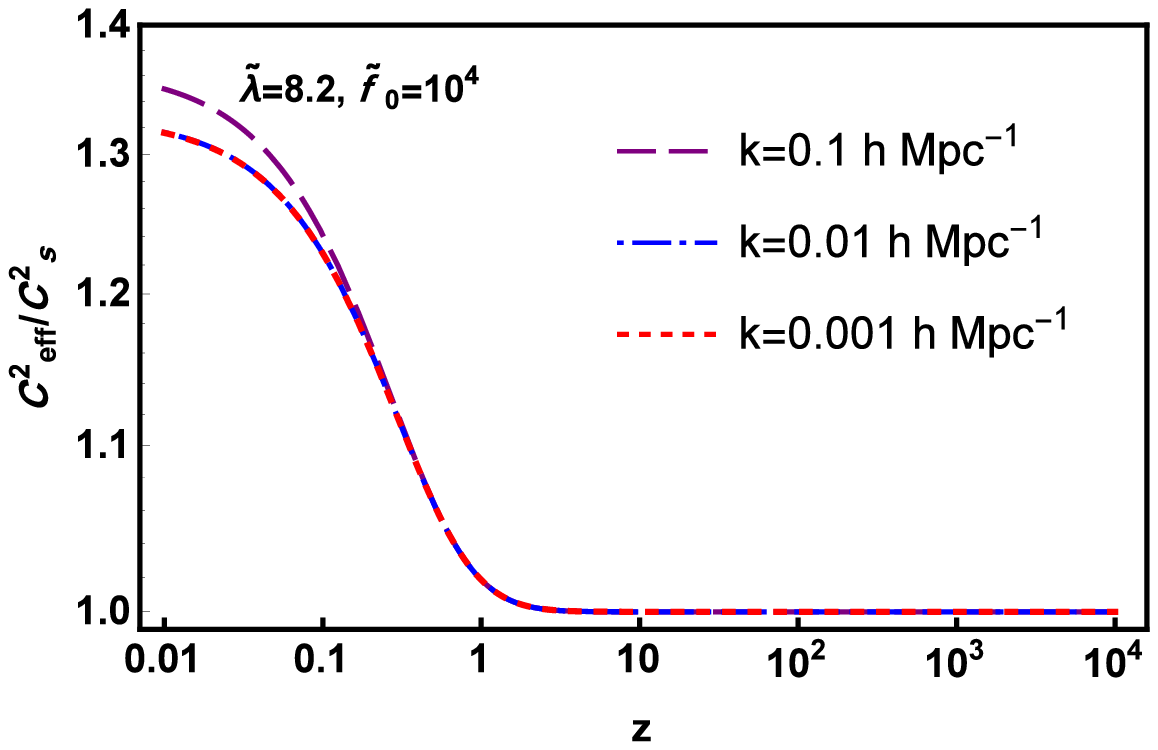}}
\end{minipage}
\caption{(a) Evolution of the scaled normalized scalar field perturbation, (b) the effective sound speed relative to adiabatic one as a function of redshift for different $k$. Auxiliary parameters are $\tilde{\lambda}=8.2$, $\tilde{f}_0=10^4$ and $\Omega_{m_0} =0.27$.}
\label{figure:varphitildek}
\end{figure}

Our analysis in the present work profits from more generality in comparison with the earlier work \cite{Fahimi2018}. Because, we solve Eqs. (\ref{dadeltam})-(\ref{lithetad}) here by using the full expression (\ref{ceff}) for the effective sound speed, whereas in the previous work \cite{Fahimi2018} the perturbation equations have been solved only in the regimes of non-clustering ($c_\mathrm{eff}=1$) and full-clustering ($c_\mathrm{eff}=0$). In the following, in Figs. \ref{phic} and \ref{Ceffc} we plot the scaled normalized scalar field perturbation $k^2\tilde{\varphi}$ and the sound speed ratio $c_{\rm eff}^2/c_s^2$ for different $k$. We see that the quantity $k^2\tilde{\varphi}$ is approximately scale independent which is in agreement with the result of Eq. (\ref{varphithetatilded}). Also the results of effective sound speed are not so sensitive to $k$ and the maximum difference is of order  $O(10^{-2})$.

\begin{figure}
\begin{minipage}[b]{1\textwidth}
\subfigure[\label{ISWa} ]{ \includegraphics[width=.48\textwidth]%
{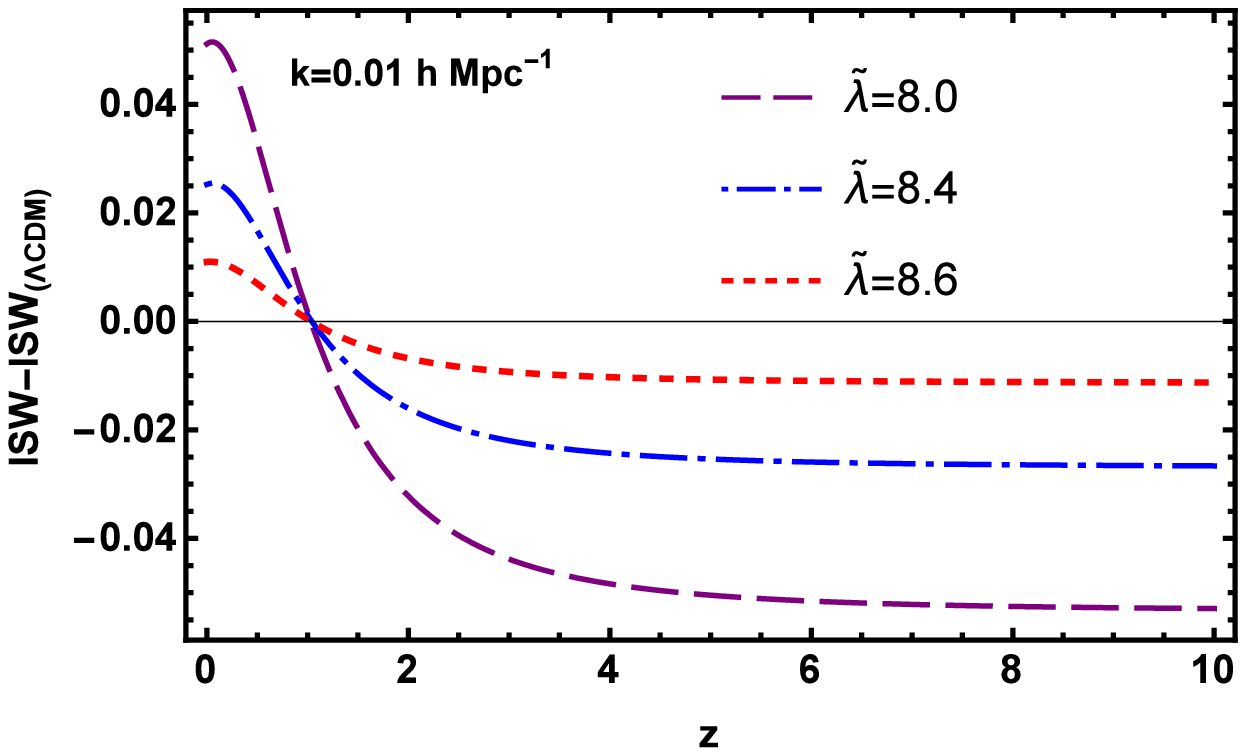}}\hspace{.1cm}
\subfigure[\label{ISWb}]{ \includegraphics[width=.48\textwidth]%
{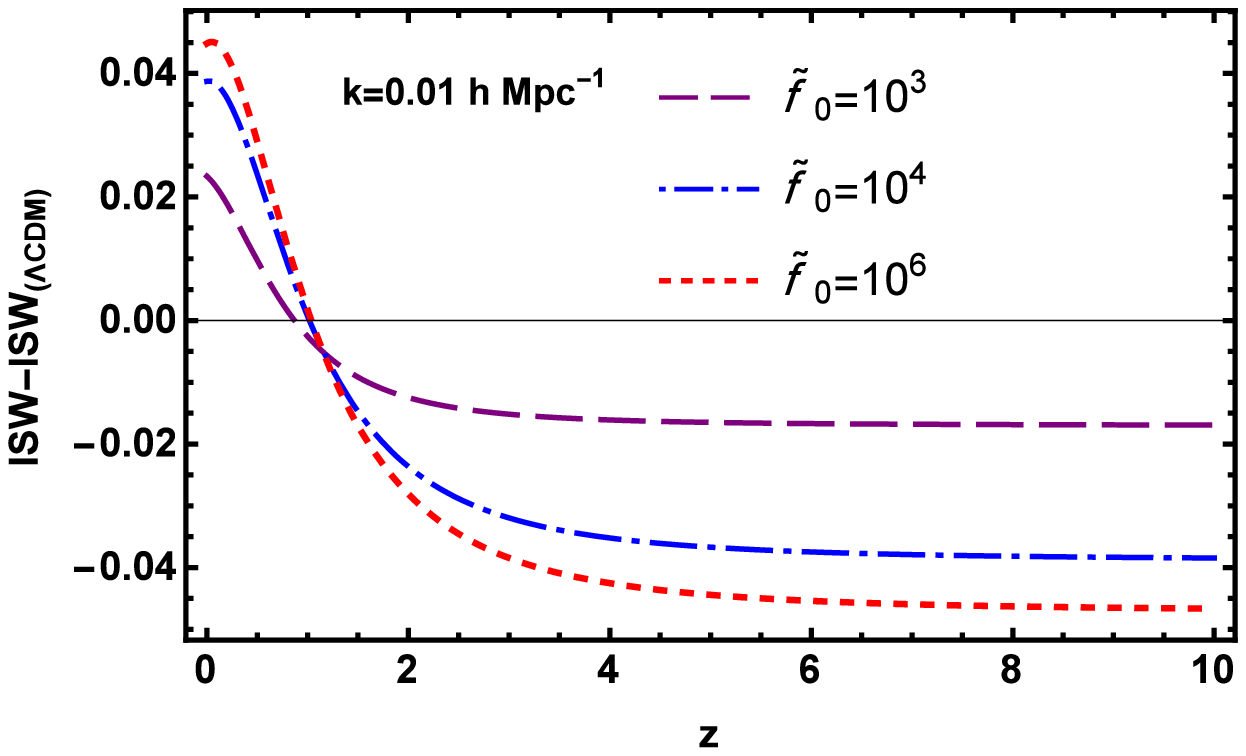}}
\subfigure[\label{ISWc}]{ \includegraphics[width=.48\textwidth]%
{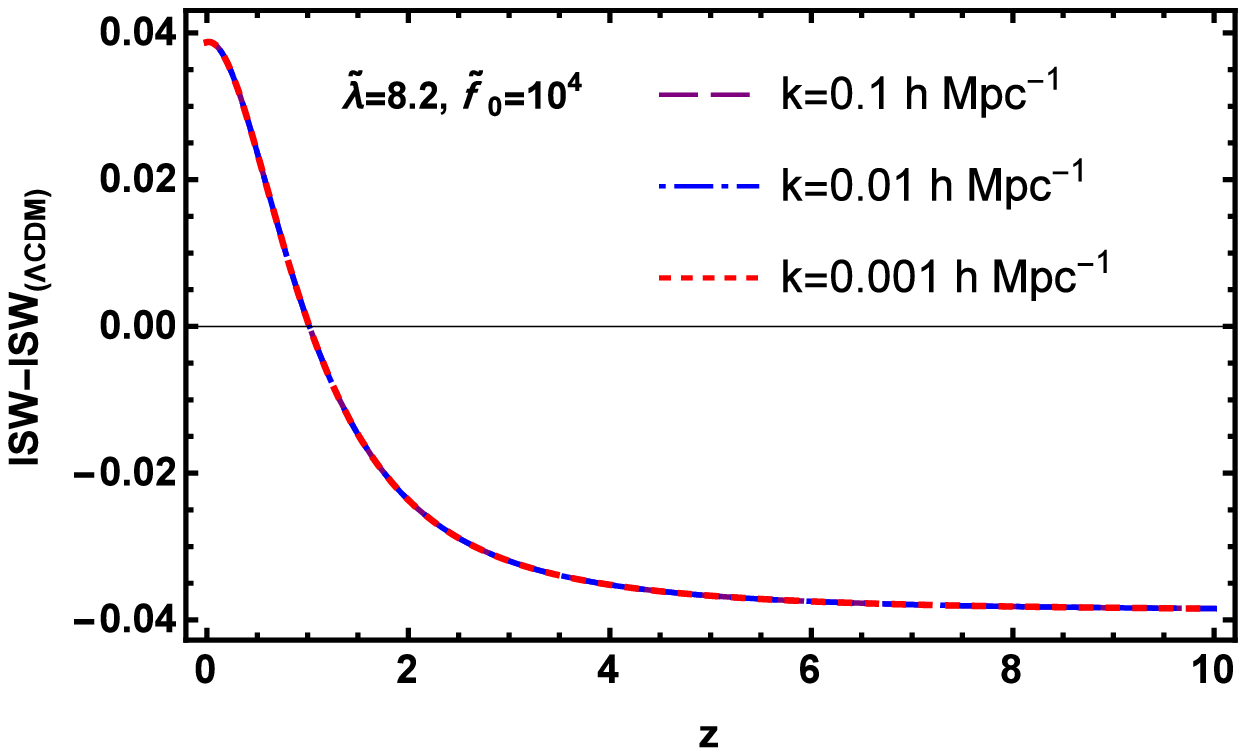}}
\end{minipage}
\caption{(a) Evolution of the ISW as a function of redshift for $\tilde{f}_0=10^5$ and different values of $\tilde{\lambda}$. (b) Same as Fig. \ref{ISWa}, but for $\tilde{\lambda}=8.2$ and different values of $\tilde{f}_0$, (c) same as Fig. \ref{ISWa} but for different wave number $k$. Auxiliary parameter is $\Omega_{m_0}=0.27$.}
\label{ISW}
\end{figure}

We can use the results of growth factor to study the evolution of the Integrated Sachs-Wolf (ISW) \cite{Sachs1967}. The ISW effect is an angular variation in the CMB temperature due to photons encountering a time-varying potential well, it means it is due to the change in the potential along the line of sight. The evolution of the relative change of the CMB temperature is given by
\begin{equation}
 \label{ISWeq}
 \tau=\frac{\Delta T}{T_{\rm CMB}}=\frac{2}{c^3} \int_{0}^{\chi_H} d \chi a^2 H(a) \frac{\partial}{\partial a} (\Phi-\Psi),
\end{equation}
where $\chi_H$ is the horizon distance \cite{Pace2013}. Also $\Phi$ and $\Psi=-\Phi$ are the gravitational potentials which are related via the Poisson equation to the matter overdensity, see Eq. (\ref{Phi}). Thus, the derivative of the potential can be related to a derivative of the matter and DE density field, effective sound speed and the scale factor. In Fig. \ref{ISW}, we plot the differences between the ISW effect of the DBI model and that obtained in $\Lambda$CDM. For all values of the model parameters, since DE perturbations affect the matter perturbations, the value of ISW for DBI models are different from the predictions in $\Lambda$CDM. As we can see at early times when the DE contribution is negligible, all the models approximate the EdS model. We will therefore have that ($\delta \propto a$), and therefore there is no contribution to the ISW effect. Also, we see that the differences between the ISW effect of the DBI model and $\Lambda$CDM are more pronounced for the smaller $\tilde{\lambda}$ or (larger $\tilde{f}_0$).

\section{Spherical collapse in the DBI model}
\label{section:spherical_collapse}

Here, we study evolution of the non-linear overdensities of DM and DE in the context of clustering DBI non-canonical scalar field. To this aim, we use the spherical collapse model (SCM) which is the simplest analytical approach to investigate structure formation in non-linear regime \cite{Gunn1972, Padmanabhan1993, Fosalba1998}. In the spherical collapse scenario, spherically symmetric regions with different peculiar expansion rates are separated from the homogeneous background. The spherical overdense regions in the presence of self-gravity interactions expand more slowly compared to the Hubble flow. This results in the density of the spherical collapsed regions enhances relative to the background content. Subsequently, the spherical overdense region tends to a maximum radius at the turnaround redshift, $z_{\rm ta}$, and detaches completely from the background content. The overdense region then starts to be collapsed independently. Finally, the spherically collapsed region is virialized at the virial redshift $z_{\rm vir}$. In the SCM, it is assumed that the density of each component of fluid follows the top-hat density profile as well as the velocity profile of each fluid is always homogeneous in the spherical region. This approach is equivalent to study the effect of perturbations to the Friedmann metric by considering spherically symmetric regions of different spatial curvature in accord with Birkhoff’s theorem \cite{Mota:2004pa}. Obviously, this model ignores any anisotropic effects of gravitational instability or collapse. It should be noted that in the accelerating Universe driven by DE, the large-scale gravitational potentials grow slower, and also the dynamical DE can be clustered and form halos in a similar way to the DM.

In the SCM, the evolution of non-linear overdensities obey the following equations \cite{Hu1998, Abramo2009, Pace2014}
\begin{align}
 & \dot{\delta}_j=-3 H (c_{{\rm eff}_j}^2-\omega_j) \delta_j-\big[1+\omega_j+\big(1+c_{{\rm eff}_j}^2\big)\delta_j\big] \frac{\theta}{a},
 \label{SC1}
 \\
 & \dot{\theta}=-H\theta -\frac{\theta^2}{3 a}-4 \pi G a \sum_j \rho_j \delta_j \big(1+3 c_{{\rm eff}_j}^2\big),
 \label{SC2}
\end{align}
where $\delta_j$, $c_{{\rm eff}_j}^2$ and $\omega_j$, are the overdensity contrast, the squared effective sound speed and the EoS parameter of $j$th component, respectively. It should be noted that, because of the top-hat density profile used in the SCM, Eq. (\ref{SC1}) holds for each fluid component $j$ separately, while Eq. (\ref{SC2}) stands alone for all the fluid components. The spherical collapse equations \eqref{SC1} and \eqref{SC2} can be rewritten in terms of scale factor in the following forms:
\begin{align}
 &\delta_{m}^{\prime}+\left(1+\delta_{m}\right)\frac{\tilde{\theta}}{a}=0,
 \label{deltamSC}
 \\
 &\delta_{d}^{\prime}+\frac{3}{a}\left(c_{{\rm eff}}^{2}-\omega_{d}\right)\delta_{d}+\left[1+\omega_{d}+\big(1+c_{{\rm eff}}^{2}\big)\delta_{d}\right]\frac{\tilde{\theta}}{a}=0,
 \label{deltadSC}
 \\
 &\tilde{\theta}^{\prime}+\left(\frac{2}{a}+\frac{H^{\prime}}{H}\right)\tilde{\theta}+\frac{\tilde{\theta}^{2}}{3a}+\frac{3}{2a}\left[\Omega_{m}\delta_{m}+\Omega_{d}\delta_{d}\big(1+3c_{{\rm eff}}^{2}\big)\right]=0.
 \label{thetasc}
\end{align}
Neglecting the non-linear terms in Eqs. (\ref{deltamSC})-(\ref{thetasc}), one gets the set of equations governing the linear overdensity evolutions as follows
\begin{align}
 &\delta_{m}^{\prime}+\frac{\tilde{\theta}}{a}=0,
 \label{scdelta}
 \\
 &\delta_{d}^{\prime}+\frac{3}{a}\left(c_{{\rm eff}}^{2}-\omega_{d}\right)\delta_{d}+\left(1+\omega_{d}\right)\frac{\tilde{\theta}}{a}=0,
 \\
 &\tilde{\theta}^{\prime}+\left(\frac{2}{a}+\frac{H^{\prime}}{H}\right)\tilde{\theta}+\frac{3}{2a}\left[\Omega_{m}\delta_{m}+\Omega_{d}\delta_{d}\big(1+3c_{{\rm eff}}^{2}\big)\right]=0.
 \label{sctheta}
\end{align}

\begin{figure}
  \centering
  \includegraphics[scale=0.9]{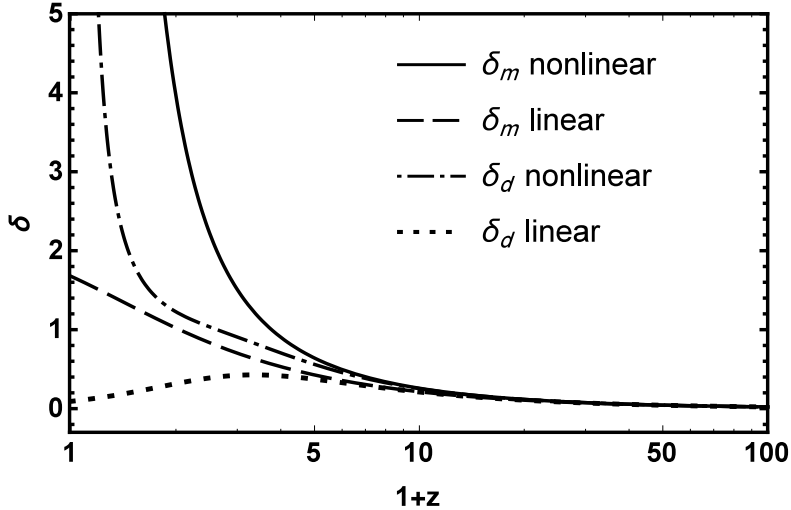}
  \caption{Non-linear and linear evolutions of $\delta(z)$ for both the DM and DBI. Auxiliary parameters are $\tilde{\lambda}_0=8.2$, $\tilde{f}_0=10^4$ and $k=0.2~\rm{h~Mpc^{-1}}$.}\label{deltanona}
\end{figure}

In Fig. \ref{deltanona}, we illustrate the evolutionary behaviours of the linear/non-linear overdensities for both the DM and DBI dark energy  for $\tilde{\lambda}=8.2$ and $\tilde{f}_0=10^4$. The figure shows that at early times the linear and non-linear perturbations overlap together, while at late times the non-linear overdensity grows very fast relative to the linear one. This result is also in agreement with that obtained by \cite{Abramo2007} for the non-phantom clustering DE (i.e. $\omega_{d}>-1$).

\begin{figure}
\begin{minipage}[b]{1\textwidth}
\subfigure[\label{varphi-non-linear} ]{ \includegraphics[width=.48\textwidth]%
{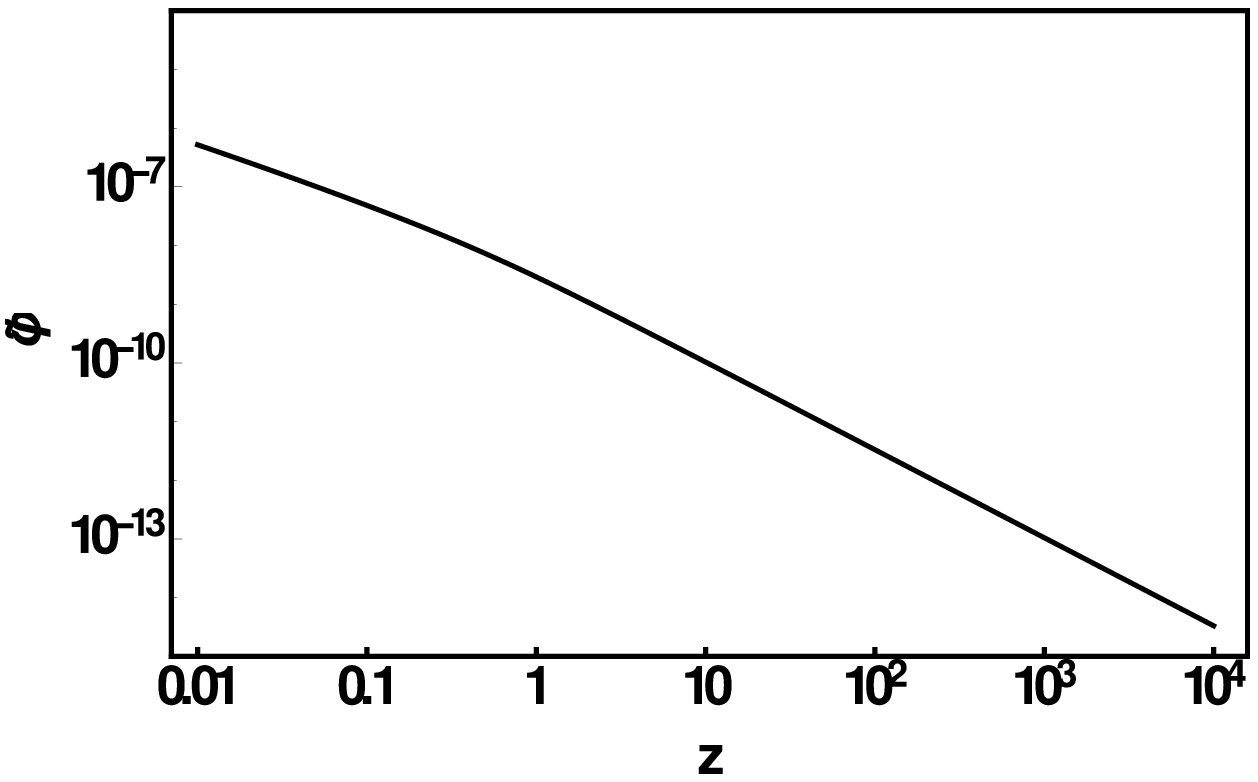}}\hspace{.1cm}
\subfigure[\label{ceff-non-linear}]{ \includegraphics[width=.48\textwidth]%
{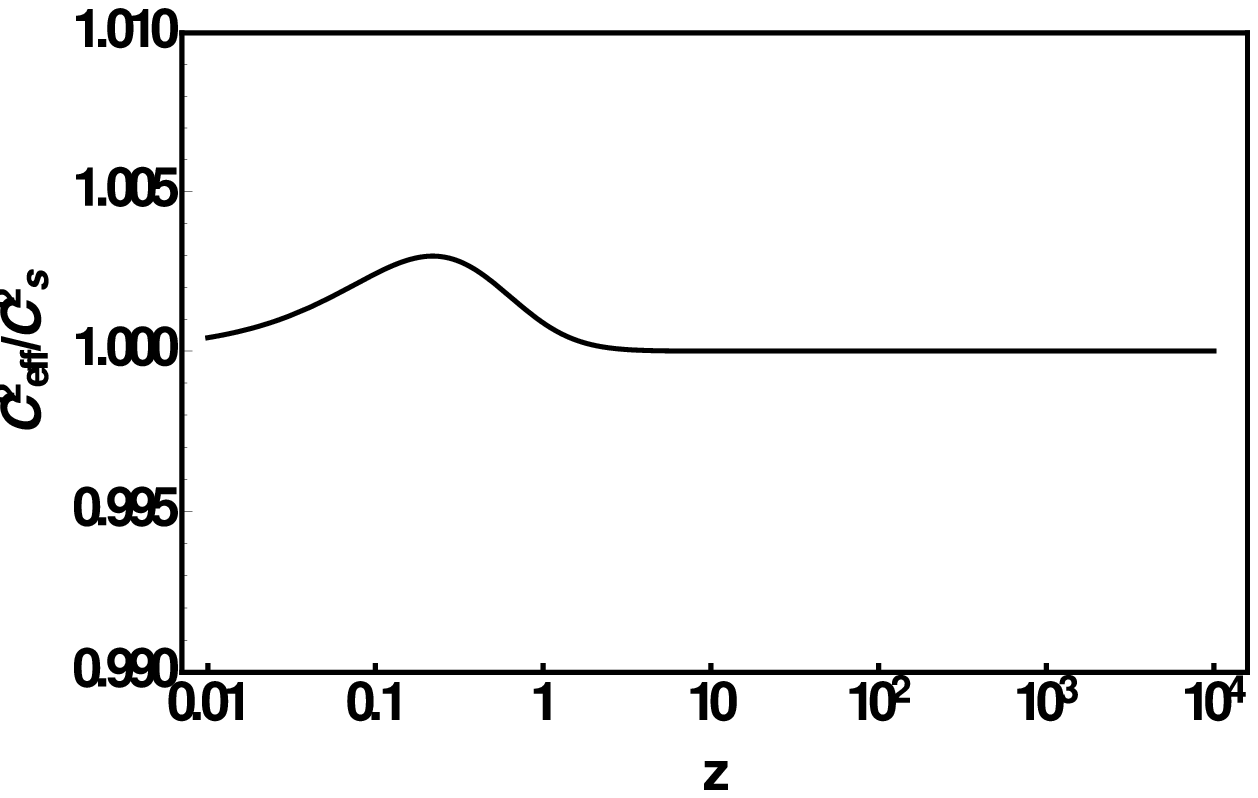}}
\end{minipage}
\caption{(a) Evolution of the normalized scalar field perturbation and (b) the effective sound speed relative to adiabatic one as a function of redshifts in non-linear regime. Auxiliary parameters are same as Fig. \ref{deltanona}.
  }\label{NonLin}
\end{figure}

Here, we should explain a point in relation to our procedure in investigating the spherical collapse formalism in our model. We know that the top-hat formulation is exact for only two limiting cases of non-clustering and full-clustering, as pointed out in \cite{Basse2011}. In the non-clustering case, the effective sound speed of DE is regarded to approach the light speed ($c_{\mathrm{eff}}=1$). In this case, the DE density contrast $\delta_d$ is effectively zero, thus the contribution of DE in the spherical collapse enters only through the background expansion. The second case is the full-clustering or comoving limit, in which the effective sound speed of DE is exactly zero ($c_{\mathrm{eff}}=0$). In this case, the Euler equations for the non-relativistic DM and DE fluids are identical, see Eq. \eqref{SC2}. This means that the bulk velocity fields of the two fluids are the same, and hence the fluids are said to be comoving. This fact however does not mean that the DE and DM density contrasts evolve identically, see Eq. \eqref{SC1}, because the conditions for energy conservation are different between the two fluids. From these points, we conclude that for the case of a variable effective sound speed $c_{\mathrm{eff}}$ the top-hat profile evolution for density perturbations is not strictly allowed and the SCM is not hold. The reason is that absence of a sharp top-hat profile leads to a scale- (or mass-) dependence in the perturbations, and this in turn means that the energy densities for both the DM and DE inside the overdense region must evolve in a non-uniform manner, even if they are initially uniform \cite{Basse2011}.

In our investigation, although we have applied the full expression of \eqref{ceff} for the effective sound speed in the spherical collapse equations \eqref{SC1} and \eqref{SC2}, the effective sound speed is very small in our work, and consequently the top-hat profile of spherical collapse formulation is applicable. To verify this, we have presented the redshift evolution of the normalized scalar field perturbation $\tilde{\varphi}$ and the effective sound speed $c_{\mathrm{eff}}$ in non-linear regime of perturbations in Figs. \ref{varphi-non-linear} and \ref{ceff-non-linear}, respectively. In Fig. \ref{varphi-non-linear}, we see that the values of $\tilde{\varphi}$ are small during evolution of the Universe. From this point and also Eq. \eqref{ceffcs}, we conclude that the values of $c_{\mathrm{eff}}$ will be very close to the values of $c_s$, as illustrated in Fig. \ref{ceff-non-linear}. It is useful to note here that due to validity of the top-hat profile of spherical collapse model in our analysis, it is allowed to consider the collapsing object as a homogeneous isotropic closed sub-universe and so the Birkhoff’s theorem holds \cite{Pace2017, Schaefer:2007nf}.


\subsection{Spherical collapse parameters}
\label{subsection:spherical_collapse_parameters}

One parameter which has an especial importance in study of SCM is the critical density or linear overdensity parameter. This quantity is defined as $\delta_c=\delta_{m \rm L}(z=z_c)$, where $\delta_{m \rm L}$ refers to linear matter density contrast, and it is computed by solving Eqs. (\ref{scdelta})-(\ref{sctheta}). For this purpose, we impose the initial conditions which lead to the divergences of the non-linear DM overdensity $\delta_m$ at a specified collapse redshift $z_c$ \cite{Pace2010, Pace2012, Pace2014}. Also, it is useful to introduce the virial overdensity which has a huge significance in spherical collapse scenario. This quantity is defined as $\Delta_{\rm vir}=\zeta(x/y)^3$, where $\zeta$ implies the overdensity at the turn around moment, $x$ indicates the normalized scale factor relative to the turn around scale factor, and $y$ refers to the ratio of the virialization radius to the turn around radius \cite{Wang1998}. It can be shown readily that in the EdS limit, $y=1/2$, $\zeta=5.6$, and $\Delta_{\rm vir}=178$, without any dependence to the redshift \cite{Meyer2012}. Note that in the presence of DE, the spherical collapse parameters can change in time. Also, as pointed in \cite{Lahav1991, Maor2005, Creminelli2010, Basse2012}, the process of virialization depends on the evolution of DE.

\begin{figure}
\centering
\includegraphics[scale=0.43]{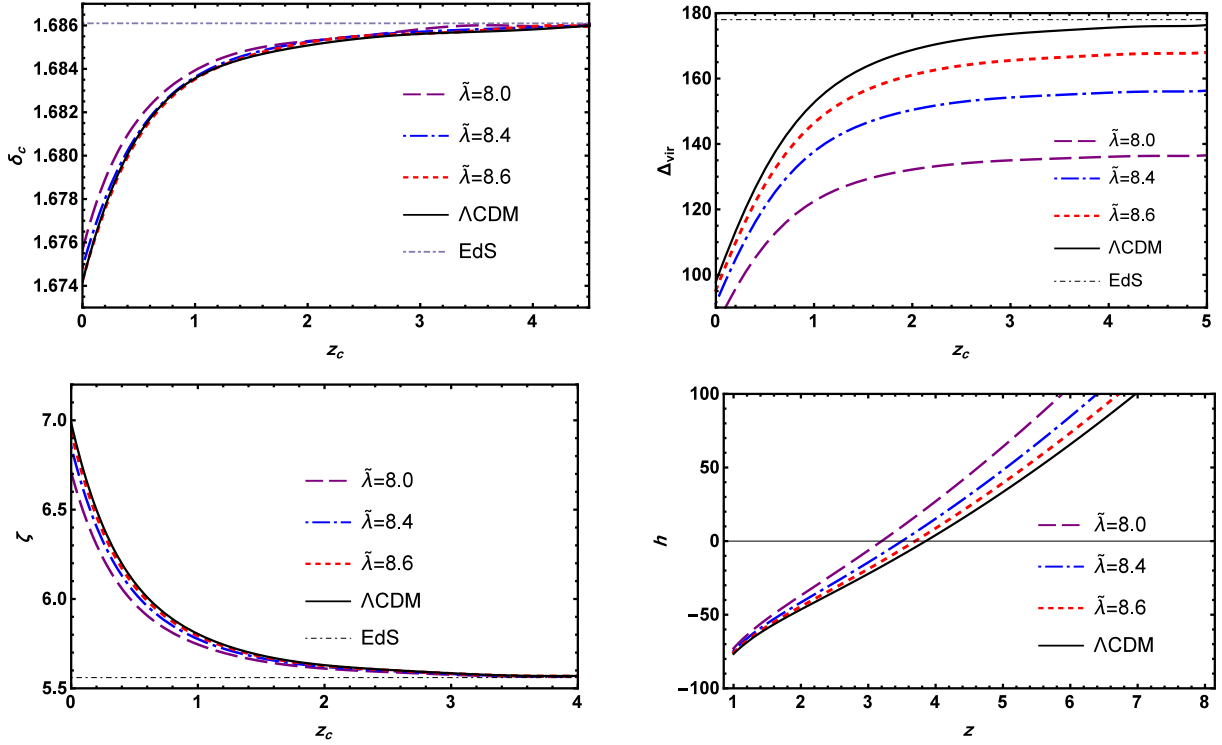}
\caption{Evolutions of the critical density contrast $\delta_c$, the virial overdensity $\Delta_{\rm vir}$, the overdensity at the turn around $\zeta$, and the rate of expansion of collapsed region $h_{\rm ta}$. Auxiliary parameters are $\Omega_{m_0}=0.27$ and $\tilde{f}_0=10^5$.}\label{figdeltac1}
\end{figure}

\begin{figure}
\centering
\includegraphics[scale=0.43]{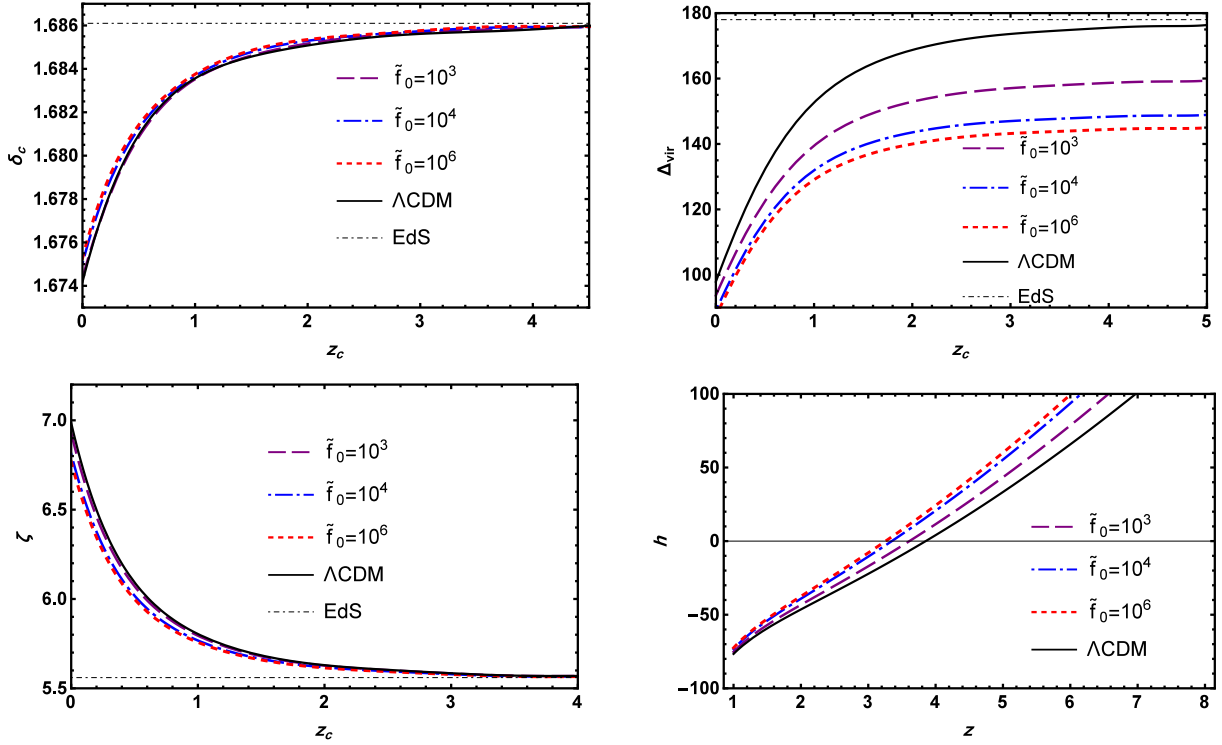}
\caption{Same as Fig. \ref{figdeltac1}, but for $\tilde{\lambda}=8.2$.}
\label{figdeltac2}
\end{figure}

In Figs. \ref{figdeltac1} and \ref{figdeltac2}, the dynamical behaviors of the linear overdensity $\delta_c(z_c)$, the virial overdensity $\Delta_{\rm vir}(z_c)$, the overdensity at the turn around $\zeta(z_c)$, and the expansion rate of the collapsed region, $h_{\rm ta}(z)= H (1+\theta/3a)$ \cite{Abramo2009JCAP}, are represented for some typical choices of model parameters. From the figures we deduce that (i) at high redshifts, the linear overdensity goes toward the expected value $\delta_c= 1.686$ in the EdS limit. It should be noted that Pace et al. \cite{Pace2017} recently showed that to satisfy the EdS limit for $\delta_c$ one should take care of in choosing both the value of the numerical infinity $\delta_{\infty}$ and the starting time for integrating the equations, $a_i$. To this aim, according to \cite{Pace2017} we take $\delta_{\infty}\geq 10^7$ and $a_i= 10^{-5}$ in our numerical computations; (ii) At lower redshifts corresponding to the DE dominated epoch, $\delta_c$ decreases and goes away from the limit of EdS universe; (iii) The virial overdensity $\Delta_{\rm vir}$ at high enough redshift tends to the EdS limit, i.e. $\Delta_{\rm vir}=178$. This is due to the fact that the pressureless dust matter is dominated and DE does not affect the structure formation. In fact, DE prevents more collapse, thus at lower redshifts when DE is dominated the density of virialized halos decreases and gets further from the limit of EdS universe; (iv) At high redshift, in our DBI model the overdensity at turn-around epoch $\zeta$ asymptotically approaches the EdS limit $\zeta=5.55$. For our DBI models, the value of $\zeta$ is smaller than one in the $\Lambda$CDM model. At lower redshifts, where the DE is dominated, the value of $\zeta$ gets further from the limit of EdS universe; (v) At the turn-around redshift $z_{\rm ta}$, the expansion rate of collapsed region $h_{\rm ta}$ transits from positive values to negative ones. For $\tilde{f}_0=10^5$ with $\tilde{\lambda} = (8,8.4,8.6)$, the turn-around redshifts are $z_{\rm ta} = ( 3.110, 3.472, 3.697)$. Besides, for $\tilde{\lambda}=8.2$ with $\tilde{f}_0= (10^3,10^4,10^6)$, the turn-around redshifts are $z_{\rm ta} = ( 3.611,3.50,3.257)$. For comparison, in  $\Lambda$CDM model the transition happens at $z_{\rm ta}=3.853$.

\subsection{Mass function and halo number density}
\label{subsection:mass_function}

Until now, we have investigated the effect of DBI dark energy perturbations on the linear overdensity threshold $\delta_c$, the virial overdensity $\Delta_{\rm vir}$, the overdensity at the turn around $\zeta$, and the expansion rate of the collapsed region, $h_{\rm ta}$. Since these spherical collapse parameters cannot be observed directly, it is more convenient to evaluate the comoving number density of virialized structures with masses in the certain range. This quantity is related closely to the observations of structure formation. Based on the Press-Schechter formalism \cite{Press1974, Bond1991}, the comoving number density of virialized structures with masses in the range $M$ and $M+{\rm d}M$ at redshift $z$ reads
\begin{equation}
 \label{dnnoncl}
 \frac{{\rm d}n (M,z)}{{\rm d} M}=-\frac{\rho_{m_0}}{M}~\frac{{\rm d}{\rm ln}\sigma (M,z)}{{\rm d}M} f(\sigma),
\end{equation}
where $\rho_{m_0}$ and $\sigma$ are respectively the present matter density and the mass fluctuation rms in spheres of mass $M$. Also  $f(\sigma)=\sqrt{\frac{2}{\pi}}\frac{\delta_c}{\sigma}\exp{\big(-\frac{\delta_c^2}{2\sigma^2}\big)}$ is the standard mass function. In spite of the fact that the standard mass function can well estimate the predicted number density of cold DM halos, but it fails by predicting too few high-mass and too many low-mass \cite{Sheth1999, Sheth2002, Lima2004}. Thus, we implement a more popular mass function given by Sheth and Tormen (ST) \cite{Sheth1999, Sheth2002} as
\begin{align}
 f_{\rm ST}(\sigma) = & A\sqrt{\frac{2 a}{\pi}}~ \left[1+\left(\frac{\sigma^2(M,z)}{a~ \delta^2_c(z)}\right)^p~\right]~ \frac{\delta_c(z)}{\sigma (M,z)}
 \nonumber
 \\
 & \times \exp\left(-\frac{a~ \delta_c^2}{2 \sigma^2(M,z)}\right),
\end{align}
with $A=0.3222$, $a=0.707$ and $p=0.3$. According to \cite{Abramo2007}, the quantity $\sigma(M,z)$ is related to its present amount as $\sigma(M,z)=D(z) \sigma_M$, in which $D(z)\equiv\delta_m(z)/\delta_m(z=0)$ is defined as the linear growth function. Moreover, $\sigma_M^2$ is the variance of smoothed linear matter density contrast given by
\begin{equation}
 \label{sigmaR}
 \sigma_M^2= \int_{0}^{\infty} \frac{d k}{k} \frac{k^3}{2 \pi ^2} P(k) W^2(kR) ~,
\end{equation}
where $R$ is the scale enveloping the mass $M=(4\pi/3)R^3 {\rho}_{m_0}$, and $W(kR)=\frac{3}{(kR)^3}\big(\sin(kR)-kR \cos(kR)\big)$ is a top-hat window function needed for the smoothing. Furthermore, $P(k)$ is the matter power spectrum of density fluctuations defined as \cite{Liddle1993, Liddle1996}
\begin{equation}
 \label{powerspectrum}
 \frac{k^3 }{2 \pi ^2}  P(k)= \delta_{H_0}^2 \left(\frac{c k}{H_0} \right)^{n_s+3} ~T^2(k)~,
\end{equation}
where $n_s=0.965 \pm 0.004$ \cite{Planck2018} is the scalar spectral index and $c$ is the light speed. Also $\delta_{H_0}$ is the normalization coefficient of the power spectrum at the present. In addition, $T(k)$ is the transfer function which depends on cosmological parameters and the matter-energy content in the Universe. In the present work, we employ the Bardeen-Bond-Kaiser-Szalay (BBKS) transfer function given by \cite{Bardeen1986},
\begin{equation}
 \label{T}
 T(x)=\frac{\ln (1+ 2.34 x)}{2.34 x}\left[1+3.89 x+(16.1 x)^2+(5.46 x)^3+(6.71 x)^4 \right]^{-1/4},
\end{equation}
where $x\equiv k/h \Gamma$ and the shape parameter $\Gamma$ is defined as \cite{Sugiyama1995}
\begin{equation}
 \label{Gamma}
 \Gamma=\Omega_{m_0} h \exp \left( -\Omega_B - \Omega_B/\Omega_{m_0} \right).
\end{equation}
In the above relation, $\Omega_B$ is the baryon density parameter, that we set it as $0.016 h^{-2}$ \cite{Copi1995, Copi1995-2}. In order to normalize the power spectrum to the same value today, we use $\sigma_{8}=\sigma_{8,\Lambda}\frac{\delta_c(z=0)}{\delta_{c,\Lambda}(z=0)}$, in which $\sigma_{8,\Lambda}=0.8120$ \cite{Planck2018} is used to normalize the matter power spectrum of $\Lambda$CDM.

\begin{figure}
\begin{minipage}[b]{1\textwidth}
\subfigure[\label{figep1}]{ \includegraphics[width=.48\textwidth]%
{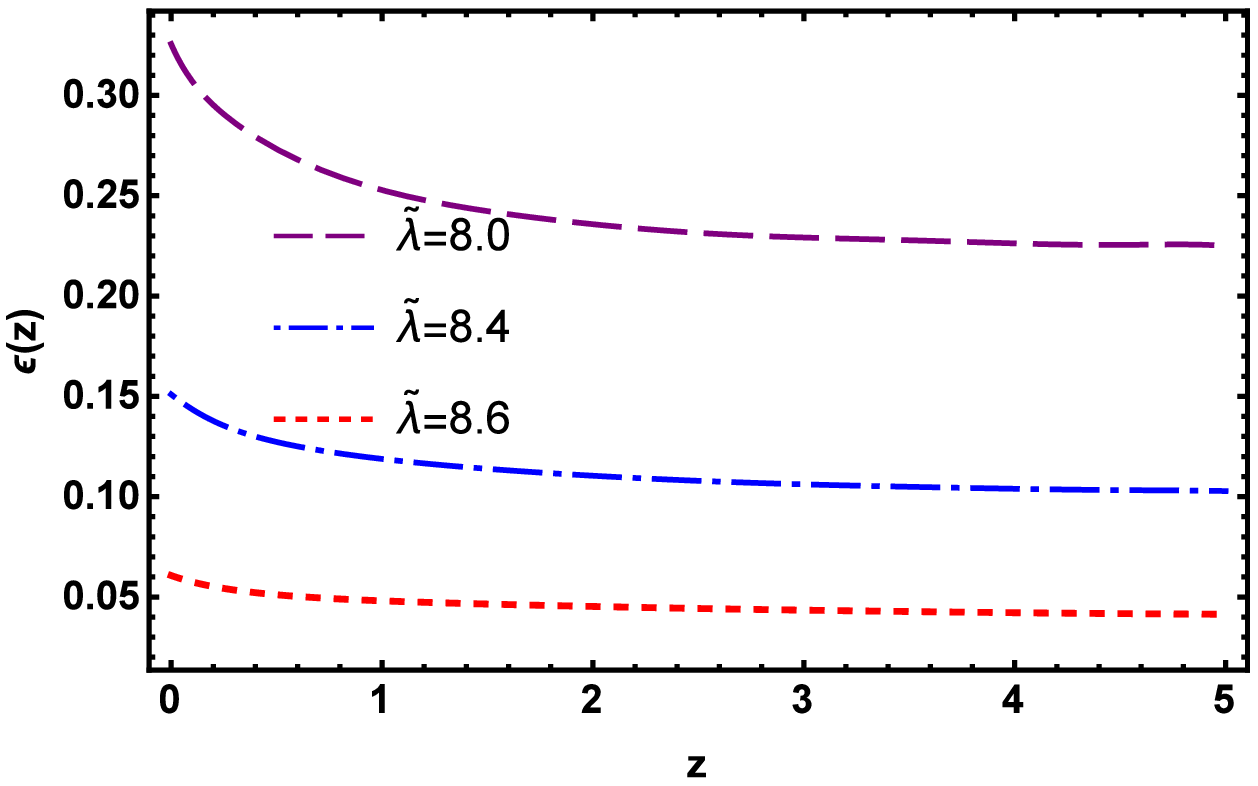}}\hspace{.1cm}
\subfigure[\label{figep2}]{ \includegraphics[width=.48\textwidth]%
{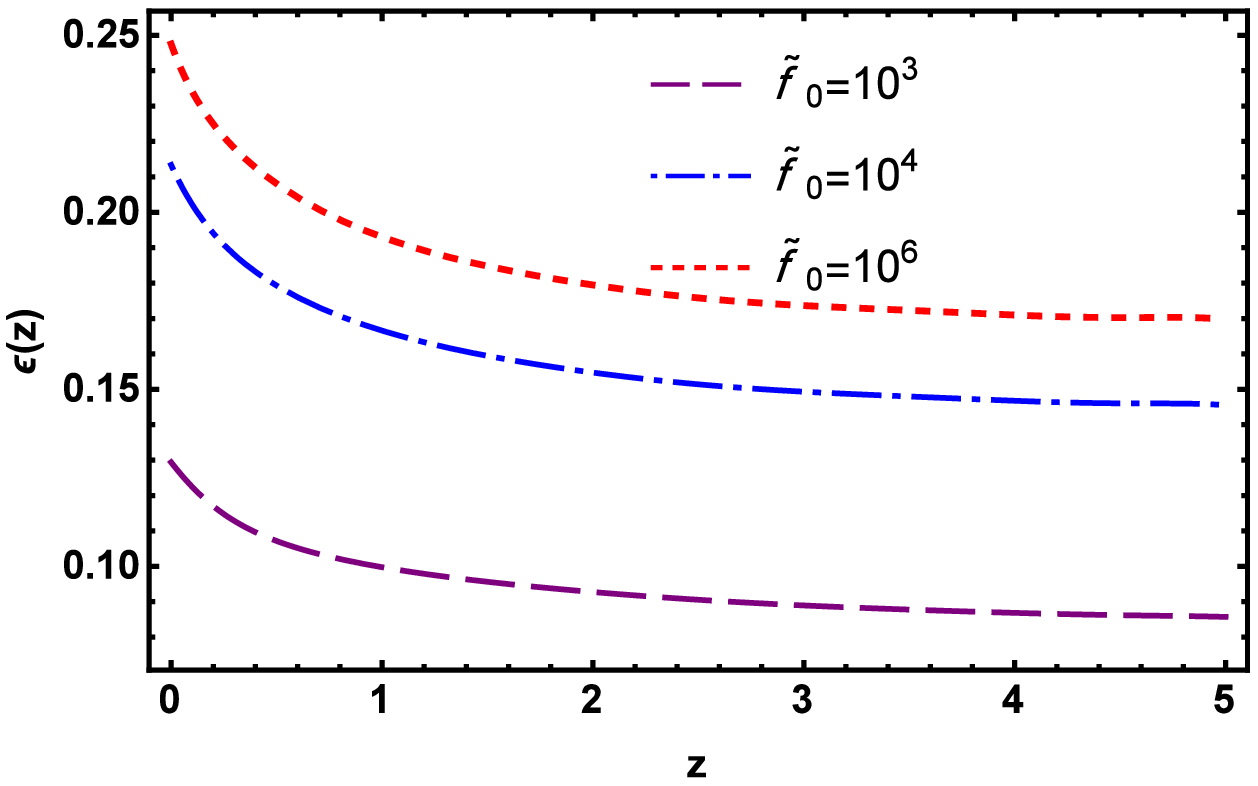}}
\end{minipage}
\caption{The ratio of DBI dark energy mass to DM mass for (a) $\tilde{f}_0=10^5$ and (b) $\tilde{\lambda}=8.2$.}
\label{figepsilon}
\end{figure}

In the presence of clustering DE, the perturbations of DE affects the halo mass \cite{Creminelli2010, Basse2011, Batista2013, Pace2014, Malekjani2015}. Consequently, in the case of clustering DE and top-hat density profile, the mass ratio of DE to DM which is defined as $\epsilon(z)\equiv M_d/M_m$ takes the form
\begin{equation}
 \label{epsilon}
 \epsilon(z)=\frac{\Omega_{d}(z)}{\Omega_m(z)}\left(\frac{\delta_{d}}{1+\delta_m}\right).
\end{equation}
In Fig. \ref{figepsilon}, we plot the variation of $\epsilon(z)$ versus the redshift for our DBI clustering DE model with different values of $\tilde{\lambda}$ and $\tilde{f}_0$. The figure implies that: (i) The DE has higher contribution to the halo mass for smaller $\tilde{\lambda}$ (larger $\tilde{f}_0$), see the left (right) panel of the figure; (ii) Because of the quintessence-like behaviour ($\omega_d>-1$) of our DBI models, we have $\delta_d>0$, and therefore according to Eq. (\ref{epsilon}), we have $\epsilon>0$ as shown in the figure.

\begin{figure}
\centering
\includegraphics[scale=0.44]{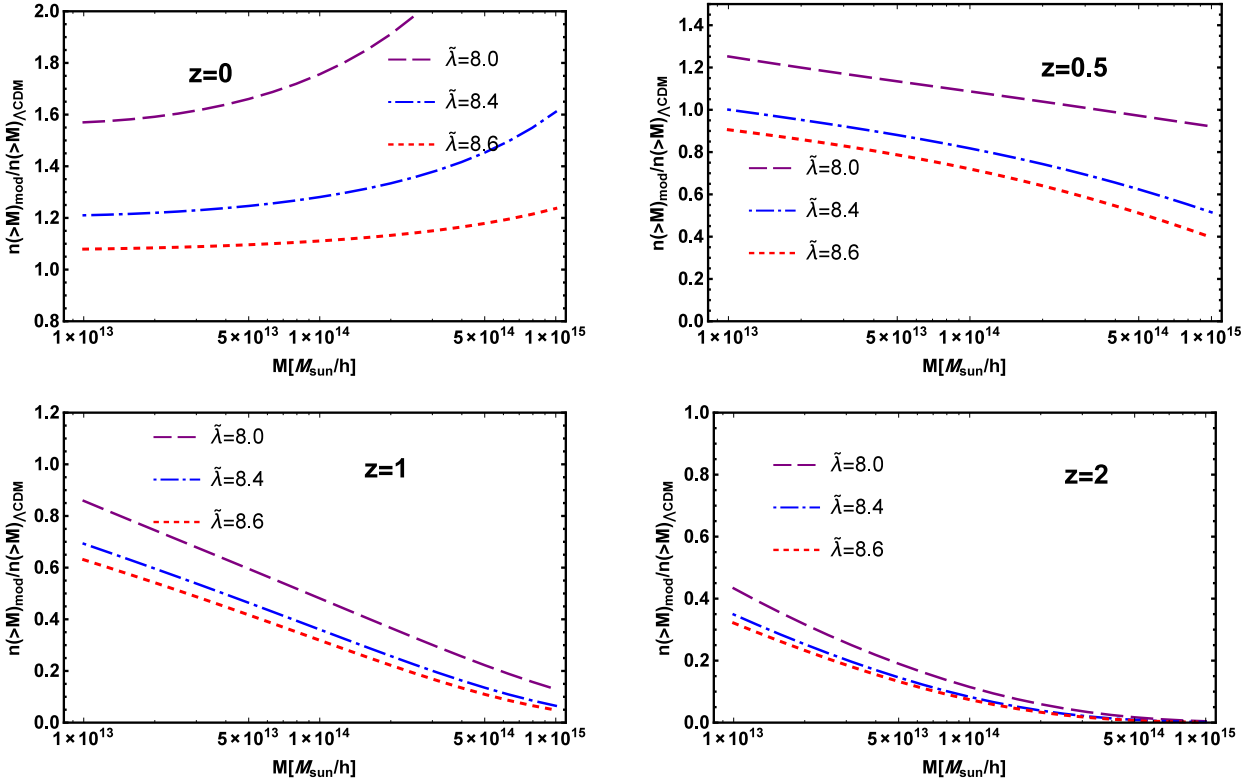}
\caption{\small{The relative number of halo objects above a given mass $M$ at the redshifts $z=0, 0.5, 1, 2$ for $\tilde{f}_0=10^5$ and different $\tilde{\lambda}$.}}
\label{mass-fig1}
\end{figure}

Taking into account the DE mass contribution to the halo mass, following \cite{Batista2013, Pace2014-2}, Eq. (\ref{dnnoncl}) is corrected as
\begin{equation}
 \label{dncl}
 \frac{{\rm d}n (M,z)}{{\rm d} M}=\frac{\rho_{m_0}}{M (1-\epsilon)}\frac{{\rm d} {\rm ln} \sigma (M,z)}{{\rm d}M} f(\sigma),
\end{equation}
where we have changed the halo mass $M$ to $M(1-\epsilon)$. Besides, the contribution of DE perturbations affects the mass function $f(\sigma)$ by changing the quantities $\delta_c$ and $\sigma(M,z)$. Here, to evaluate the number density of halo objects above a given mass at fixed redshift as $n(>M)=\int_{M}^{\infty} \frac{{\rm d} n}{{\rm d}M'}~ {\rm d}M'$, we use Eqs. (\ref{dncl}), for our clustering DBI non-canonical scalar field models of DE.
We plot the relative number density of halo objects above a given mass at redshifts $z=0, 0.5, 1, 2$ for our DBI scenarios in Fig. \ref{mass-fig1}. At the present time, $z=0$, we observe that DBI model predicts
more abundance of virialized halos than the $\Lambda$CDM cosmology at both the
low and high mass tails. We see that at higher redshifts, differences between various models are so small. Comparing different pannels of Fig. \ref{mass-fig1} shows that the number density of halos decreases with increasing the redshift. As expected, this decrement is more pronounced for massive halos compare to low mass halos. This result is compatible with this fact that the low mass halos are formed before larger ones.

\begin{figure}
\centering
\includegraphics[scale=0.44]{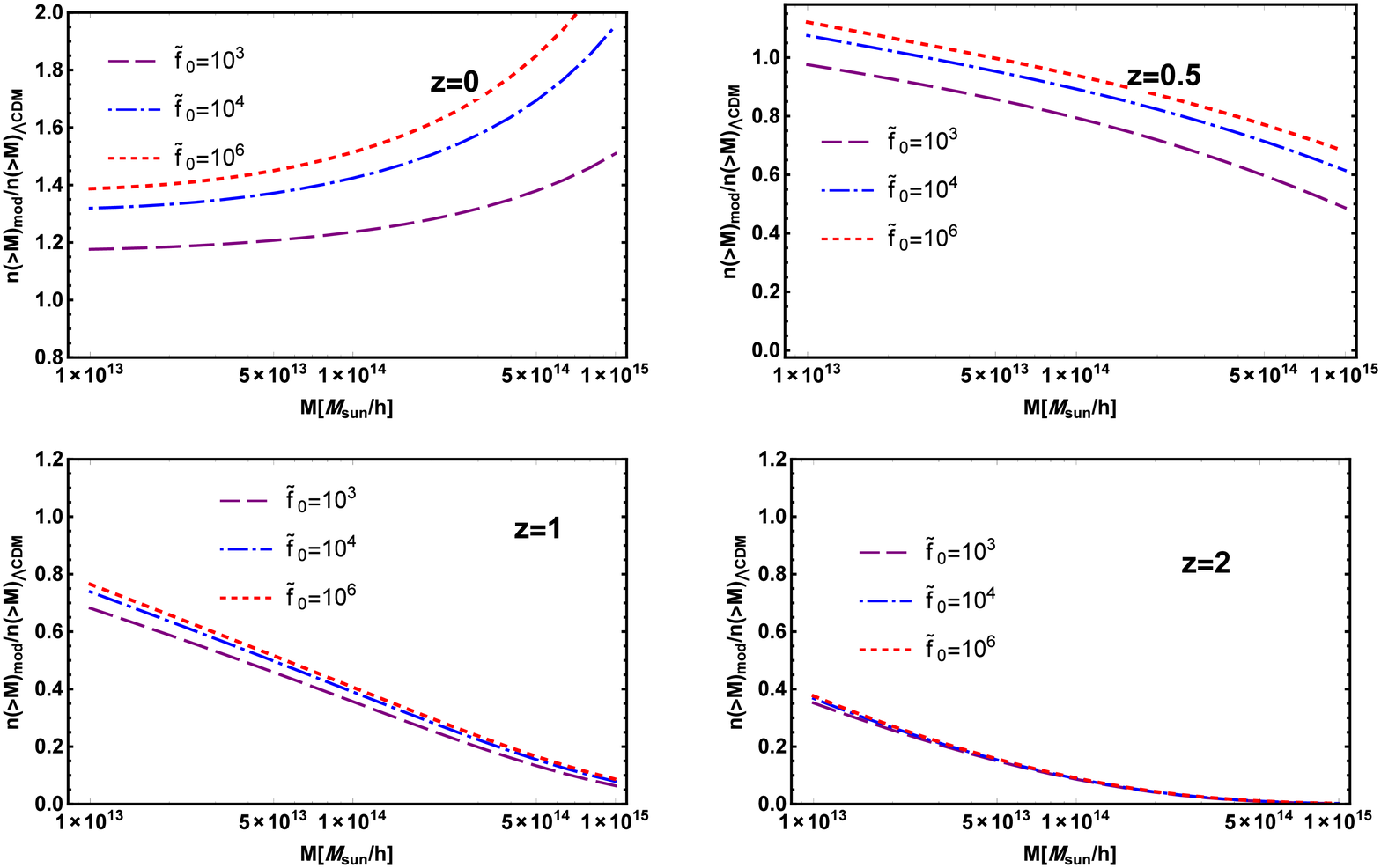}
\caption{\small{Same as Fig. \ref{mass-fig1}, but for $\tilde{\lambda}=8.2$ and different $\tilde{f}_0$.}}
\label{mass-fig2}
\end{figure}

In Fig. \ref{mass-fig2}, we illustrate the results similar to Fig. \ref{mass-fig1} but for different $\tilde{f}_0$. 

\section{Conclusions}
\label{section:conclusions}

Within the framework of DBI non-canonical scalar field models of DE, we investigated the perturbations of DM and DE in both the linear and non-linear regimes. The setup of our DBI model is characterized by the AdS warp factor $f(\phi)=f_0/\phi^4$ and the quartic potential $V(\phi)=\lambda \phi^4/4$. At the background level and in a flat FRW universe, we obtained the evolutionary behaviours of the normalized Hubble parameter $\tilde{H}$, normalized DBI scalar field $\tilde{\phi}$, density parameters ($\Omega_m$, $\Omega_d$), deceleration parameter $q$, EoS parameter $\omega_d$, and effective EoS parameter $\omega_{\rm eff}$. We concluded that: (i) in our DBI model the background quantities for smaller $\tilde{\lambda}$ (or larger $\tilde{f}_0$) more deviate from the $\Lambda$CDM model; (ii) The EoS parameter of DBI DE $\omega_d$ has a quintessence-like behaviour, i.e. $\omega_d>-1$; (iii) The effective EoS $\omega_{\rm eff}$ and deceleration $q$ parameters start from a matter-dominated universe $(\omega_{\rm eff}=0,q=0.5)$, as expected, and then tend to the EdS limit $(\omega_{\rm eff}=-1,q=-1)$; (iv) In our DBI model, the transition redshift $z_t$ from deceleration ($q>0$) to acceleration ($q<0$) era is nearly close to that of the $\Lambda$CDM model.

In our analysis, we calculated the full expression for the effective sound speed $c_{\mathrm{eff}}$. As a result, this quantity can vary with time in our investigation during the cosmological evolution. Besides in our work, we considered a variable sound speed. Hence, our analysis in the present work is more general and realistic than the our previous investigation \cite{Fahimi2018} and also relative to most of the previous studies which are limited to only the two cases of non-clustering ($c_{\mathrm{eff}}=1$) and full-clustering ($c_{\mathrm{eff}}=0$) scenarios. With the help of PN formalism, we evaluated the linear growth rate of DM, $D=\delta_m/\delta_{m_0}$, in our clustering DBI setting. Our analysis implies that the perturbation of DBI scalar field begins from small values in the initial times, and then it grows very slowly. Therefore, value of the scalar field perturbation remains very small during all history of the Universe. As a result, the effective sound speed of the perturbations is approximately equal to the adiabatic sound speed of the model. We pointed out that the result of growth factor in the DBI models is smaller than $\Lambda$CDM model, and although $k$ is appeared in our calculation, but $D/a$ is approximately scale independent. We also estimated the ISW parameter in our setup, and showed that our DBI dark energy model manifests some deviation from the concordance $\Lambda$CDM model. This fact may be regarded in the future to discriminate between our dynamical DE and the $\Lambda$CDM model in light of the CMB observational data.

In the next step, we examined the non-linear growth of DM and DBI perturbations. For this purpose, since in our model we concluded that $c_{\mathrm{eff}} \approx c_s \ll 1$, hence we applied the top-hat profile based on the spherical collapse formulation. We computed the spherical collapse parameters including the linear overdensity $\delta_c(z_c)$, the virial overdensity $\Delta_{\rm vir}(z_c)$, the overdensity at the turn around $\zeta(z_c)$ and the rate of expansion of collapsed region $h_{\rm ta}(z)$. Our results show the following. (i) The linear overdensity $\delta_c$ at high redshifts, goes toward the expected value $\delta_c = 1.686$ in the EdS limit; (ii) The virial overdensity $\Delta_{\rm vir}$ and the overdensity at the turn around $\zeta$ in initial times converge to $178$ and $5.55$, respectively, which are the same values corresponding to the EdS universe. This happens because the impact of DE at high redshifts is negligible; (iii) As parameter $\tilde{\lambda}$ ($\tilde{f}_0$) increases (decreases), the expansion rate of the collapsed region $h_{\rm ta}(z)$ behaves like that in the $\Lambda$CDM model. In addition, $h_{\rm ta}$ alters its sign at higher redshifts in $\Lambda$CDM model  in comparison with the DBI models. In other words, the turn-around takes places earlier in the $\Lambda$CDM model than DBI models. The parameters of the SCM make it also possible for us to estimate the fraction of DE mass to the DM mass of DM, $\epsilon(z)=M_d/M_m$, and rational number density of the virialized halos exceeding an arbitrary mass, $\frac{n(>M)_{\rm DBI}}{n(>M)_{\Lambda \rm CDM}}$.  Our results imply that for $z=0$, the rational number density of the objects in the DBI models is more than the $\Lambda$CDM result. Besides, for higher redshifts the abundance of halos falls down for massive halos. Because the low mass DM halos form sooner than the larger one.

\section*{Acknowledgments}

The authors thank the anonymous referee for very valuable comments.













%


\end{document}